\documentclass[journal, comsoc]{IEEEtran}

\usepackage{graphicx}
\usepackage{amssymb}
\usepackage{amsmath}
\usepackage[hyphens]{url}
\usepackage{amsfonts}
\usepackage{cite}
\usepackage{url}
\usepackage{enumerate}
\usepackage{epstopdf}
\usepackage{dsfont}
\usepackage{steinmetz}
\usepackage{amsthm}
\usepackage{balance}
\usepackage[mathscr]{euscript}
\usepackage{color}

\newtheorem{proposition}{Proposition}

\allowdisplaybreaks[1]

\DeclareSymbolFont{sfletters}{OT1}{cmss}{m}{n}
\DeclareMathSymbol{\sGamma}{\mathord}{sfletters}{"00}
\DeclareMathSymbol{\sDelta}{\mathord}{sfletters}{"01}

\makeatletter
\def\footnoterule{\relax%
  \kern-5pt
  \hbox to \columnwidth{\hfill\vrule width \columnwidth height 0.4pt\hfill}
  \kern4.6pt}
\makeatother

\begin{document}
\title{Sensitive and Nonlinear Far Field RF Energy Harvesting in Wireless Communications}

\author{Panos~N.~Alevizos,~\IEEEmembership{Student Member,~IEEE} 
	  and Aggelos~Bletsas,~\IEEEmembership{Senior Member,~IEEE}
\vspace{-0.2 in}
\thanks{This research is implemented through the Operational Program ``Human Resources Development, 
Education and Lifelong Learning'' and is co-financed by the European Union (European Social Fund) and Greek national funds.

Authors are with  School of Electrical and Computer Engineering (ECE), Technical University 
of Crete, Chania 73100, Greece. (E-mail: palevizos$@$isc.tuc.gr, aggelos$@$telecom.tuc.gr) 
}}


\maketitle

\begin{abstract}
This work studies \emph{both} limited sensitivity 
\emph{and} nonlinearity of far field RF energy harvesting observed in reality and quantifies their effect,
attempting to fill a major hole in the simultaneous wireless information and power transfer (SWIPT) literature. 
RF harvested power is modeled as an arbitrary nonlinear, continuous, and non-decreasing function of received power,
taking into account limited sensitivity and saturation effects. RF harvester's sensitivity may be several dBs 
worse than communications receiver's sensitivity, potentially rendering RF information signals
useless for energy harvesting purposes. Given finite number of datapoint pairs of harvested (output) power
and corresponding input power, a piecewise linear approximation is applied and the statistics of the harvested 
power are offered, as a function of the wireless channel fading statistics.  Limited number of datapoints are needed 
and accuracy analysis is also provided. Case studies include duty-cycled (non-continuous), as well as continuous SWIPT, 
comparing with industry-level, RF harvesting. The proposed approximation, even though simple, offers
\emph{accurate performance} for all studied metrics. On the other hand, linear models or nonlinear-unlimited
sensitivity harvesting models deviate from reality, especially in the low-input-power regime. 
The proposed methodology can be utilized in current and future SWIPT research.
\end{abstract}


\IEEEpeerreviewmaketitle

\section{Introduction}
%




Far field radio frequency (RF) energy harvesting, i.e., the capability
of  wireless nodes to scavenge energy, either from remote ambient or
dedicated RF sources, has  recently attracted significant
attention. Compared to other energy harvesting methods, e.g., from motion, sun or heat, RF energy harvesting offers the advantage of 
simultaneous wireless information and power transfer (SWIPT). The latter lies at the heart of the radio frequency identification (RFID) industry,
  which is expected to drive research and innovation in a plethora of coming Internet-of-Things (IoT) scenarios and low-power applications \cite{Alevizos_PhD_thesis:17}.

  Recent SWIPT   literature within the  wireless communications theory research community has addressed   problems relevant to    protocol architecture,
 as well as fundamental performance metrics.   Several motivating examples demonstrating the   concept  of SWIPT exist in the literature, e.g., 
 for  memoryless point-to-point channels  \cite{Var:08}, 
 frequency-selective channels \cite{GroSah:10}, 
 multiple-input multiple-output (MIMO) broadcasting \cite{ZhaHo:13}, and relaying \cite{NaZhDuKe:13}.  
 For instance, work in \cite{NaZhDuKe:13} studied protocols that split time or power
 among   the RF energy harvesting and    information transfer modules within a radio terminal, so that specific communication tasks are
 performed, while the radio terminal is solely powered by {the} receiving RF. Wireless power transfer in wireless communications imposes
 additional energy harvesting constraints \cite{ZhZhCh:13_b}. 
  Work in \cite{HuLa:13} offered 
 several resource allocation algorithms  for wideband RF harvesting systems. 
  The reviews in   \cite{KrTimNikZhNgWiSch:14, UlYeEkSiZoGrHu:15}
 offer the current perspective of linear RF harvesting within   the  wireless communications theory community.

On the other hand, RF energy harvesting suffers from limited available density issues, typically in the sub-microWatt regime 
(e.g., work in \cite{TEX:10} reports $0.1 \mu$Watt/cm$^2$ from cellular GSM base stations), in sharp contrast to other ambient
energy sources based on sun, motion or electrochemistry;\footnote{For example, sun can offer $35$mW/cm$^2$ using a low-cost
5.4cm $\times$ 4.3cm polycrystalline blue solar cell \cite{FUT:15}, while electric potential across the stem of a 60 cm-tall 
avocado plant can offer $1.15 \mu$Watt at noon time \cite{KonkouMitBl:16}.} such limited RF density can power only  ultra-low-power 
devices in continuous (non-duty-cycled) operation or  low-power devices, such as   low-power wireless sensors in delay-limited, 
duty-cycled operation, since sufficient RF energy must be harvested   before operation.   That is due to the fact that the far field
  RF power decreases  at least quadratically  with distance,
  while RF harvesting circuits have 
  \emph{limited} sensitivity,  i.e., offer no output when input power is below a threshold,  as well as efficiency.  
  A common, 
  critical component of  the far field RF harvesting circuits is the rectenna, i.e., 
the antenna and the rectifier that converts the input RF signal to DC voltage.

\begin{figure}[!t]
 \centering
         \includegraphics[scale=0.57]{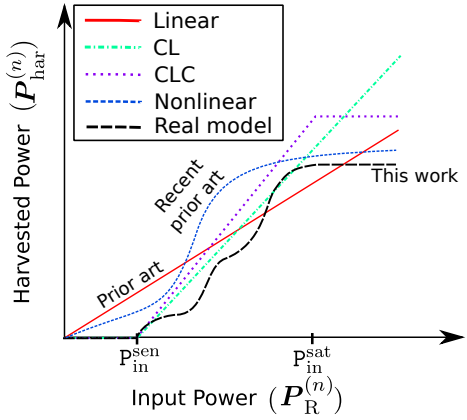}
  \caption{Harvested power vs. input  power.
  For the real rectenna model, the harvested power is an increasing function of
    input  power, taking into account the effect of   harvester's   sensitivity.}
 \label{fig:problem}
 \end{figure}

%
The rectifier circuit
is typically implemented with one or multiple diodes, imposing strong nonlinearity
on the power conversion. 
In addition, the rectifier circuit has usually three operation regimes, stemming directly 
from the presence of diodes.
First, for input power below  the \emph{sensitivity}  of the harvester (i.e., the minimum power for harvesting operation), 
the harvested power is zero.  Second, for input power between sensitivity and saturation  threshold (the power level above which the
output harvesting power saturates), the harvested power is a continuous, nonlinear, increasing function
of   input RF power, with response depending on the operating frequency and the circuit components
of the rectifier. Lastly, for input power  above saturation, 
the output power of the harvester is saturated. The above three characteristic regimes are depicted
in Fig.~\ref{fig:problem}, with the black-dashed line curve,
{which} adhere to a variety of circuits in the microwave literature \cite{ViVu:13, VaDu:14, PopFaErCoZa:13, AsDaBle:16, PowerCast}. 
The nonlinearity of harvested power as a function of input power is also corroborated by the fact that the conversion 
\emph{efficiency} in the microwave circuits literature is always referenced to a specific level of input power. 
%

 %

 There exist few recent   SWIPT reports  studying nonlinear RF harvesting models, i.e., modeling the harvested 
 power as a specific nonlinear function of the input power. Modeling of the harvesting power as a normalized sigmoid function is proposed in 
 \cite{BoNgZlSc:15,  BoNgZlSc:17_a,  BoNgZlSc:17_b, BoNgZlSc:17_c, BoNgZlSc:17_d},
whereas work in \cite{XuOzAyMcKVi:17} models the harvested power as a second order polynomial. 
 These studies examine resource allocation algorithms under nonlinear RF harvesting using convex optimization techniques;
however, the adopted nonlinear RF harvested power models do not  account for the harvester's limited sensitivity, i.e., 
sensitivity threshold is assumed zero and the harvester can output power for any non-negative input power value.  

 There is an important difference between the \emph{communications receiver's sensitivity} 
and the harvester's sensitivity (defined above), largely overlooked   by a wide portion of SWIPT prior art in wireless communications.
The first one is the minimum power threshold   above  which the receiver can   reliably 
decode signals, with values that depend on the temperature, bandwidth, noise figure of the electronics and the minimum 
signal-to-noise ratio (SNR). Communication sensitivity ranges  from  $-140$ dBm (e.g., for low-bandwidth radios such as 
LoRa \cite{Tal_et_al:17}) to $-85$ dBm (e.g., higher bandwidth GSM cellphones). 
On the other hand,  the state-of-the-art harvesting sensitivity currently obtains values in the order of $-30$  dBm; unfortunately, the harvester's sensitivity  
evolves very slowly as a function  of years  (slower than Moore's law), due to the involved semiconductor technology; e.g., passive RFID tags harvester sensitivity  
(in dBm)  has improved by a factor of two every $3.8$
years  over  a  two-decade  span  \cite[Fig.~1]{Du:16}.   As a result,
there is a non-negligible gap around 
$55-120$ dB between communications receiver's and harvester's sensitivity. This gap
indicates that the signals with power around communications sensitivity 
 can be decoded at a SWIPT receiver but cannot be exploited for energy harvesting purposes.

Work in \cite{TroGriDur:09} proposed exploitation of peak-to-average power ratio (PAPR), when the power of the emitted signal
is spread over multiple tones; the peaky behavior of the multi-tone emitted signal can offer adequate bursts  of energy to
the rectifier, turning on the diode,
even if the average input power is below the harvesters' sensitivity.
Prominent signal examples are multi-sine waveforms \cite{BoCa:11} or orthogonal frequency-division multiplexing (OFDM) waveforms.  
Subsequent work \cite{HuCl:16,  Cler:18,   ZheClZh:17, HuCl:17_a, VarRasCle:17} optimized  amplitudes and phases 
of the multi-tone waveforms, maximizing the harvested power at the receiver, under flat or frequency-selective channels.
Convex optimization techniques were employed, with   
channel state information (CSI) at the transmitter, PAPR  constraints and nonlinear, input-output circuit-based 
analysis of a single-diode or multiple-diode rectifiers \cite{CleBaz:16, CleBaz:17}.  Experimental measurements \cite{KimCleMit:17} 
demonstrated that the harvesting efficiency of   multi-tone systems can be increased by $37\%$
 compared to   single-tone, within the low-input-power range of $[-28,-19]$ dBm. Although the PAPR
property of multi-tone signals can increase the end-to-end harvesting efficiency, the level of the studied input powers
 was still above  $-30$ dBm, while the state-of-the-art RF harvesting sensitivity is currently close to  $-35$ dBm \cite{AsDaBle:16}. 
 More importantly, the effect of limited RF harvesting sensitivity has not been quantified in the context of SWIPT research.
  

  Therefore, the majority of SWIPT studies within the wireless communications community,
to the best of our knowledge, either (a) adhere to a  linear model of harvested power as a
function of input RF power \emph{or} (b) do not take explicitly into account the effects of 
harvester's limited (and not unlimited) sensitivity; the latter is of vital importance, given the
fluctuations of received signal input power due to wireless fading, as well as the fact that the harvester's 
sensitivity is finite and several tens of dB worse than communications receiver's sensitivity.

This work  introduces \emph{both} limited sensitivity \emph{and} nonlinearity of far field RF energy harvesting 
observed in reality, attempting to fill a major hole in the SWIPT wireless communications theory community.
Two rectifier circuit harvesting efficiency models are examined from the prior art 
 for realistic comparison; the first one is the  sensitive rectenna proposed in
\cite{AsDaBle:16} and the second is the   PowerCast module \cite{PowerCast}.
Three (approximation) baseline harvested power  models are compared with 
the  realistic  harvested power model,  depicted in Fig.~\ref{fig:problem}. The first baseline model called linear (L),
is the  dominant model of RF harvesting prior art.
The other two studied baseline models are called constant-linear (CL)
and constant-linear-constant (CLC).  
Additionally,   nonlinear harvesting models with unlimited sensitivity are also studied and compared with the approach of this work. The contributions are summarized below:
\begin{itemize}
 \item For the first time in the literature,  harvested power can be
 modeled as an arbitrary nonlinear, continuous, and non-decreasing function of the input RF power, 
 taking into account (a) the nonlinear efficiency of realistic rectifier RF harvesting circuits,
 (b) the zero response of energy harvesting circuit for input power below sensitivity (i.e., limited sensitivity), and (c) the saturation effect
 of harvested power.%
 \footnote{ 
Harvester's saturation power levels obtain nominal values on the order of several tens of milli-Watts; such numbers 
are not often encountered in practice, since they imply short transmitter-receiver distance or very large transmission
power. 
However, saturation threshold effect exists in any
RF harvesting circuitry due to the presence of diode(s) \cite[Fig. 3]{VaDu:14}.
As discussed in \cite[Remark 5]{Cler:18}, the saturation effect 
can be avoided in the input range of interest by properly designing the rectifier.
For ultra-small-range applications, as in specific RFID systems,
there is possibility for the RF harvester to operate close or above the saturation  threshold.
}
  The  impact of harvester's  limited  sensitivity is carefully quantified based on the characteristics 
 of the RF harvesting circuitry and the wireless propagation channel. 
\item   Given the wireless channel fading probability density function (PDF) and datapoint pairs of the harvested (output) power and  the corresponding 
input power, stemming from the specifications of the limited-sensitivity, nonlinear 
harvesting system, this work offers the PDF and cumulative distribution function (CDF) of the harvested  power. The offered statistics are based on a piecewise
linear approximation. It is also shown that approximation accuracy of at least $\epsilon$ can be achieved by at most $\mathcal{O}(\sqrt{1/\epsilon})$ datapoints.     
 \item Three performance metrics are studied: (i) the expected harvested energy at the receiver, 
  (ii) the expected charging time at the receiver (time-switching scenario), and (iii) the probability of
  successful reception at  the interrogator for passive RFID tags (power-splitting scenario).
    It is shown that the proposed  approximation  methodology
  offers \emph{exact performance} for all studied
  metrics.  In addition, no tuning of any parameter is required. On the other hand,
  linear RF harvesting modeling results deviate from reality, and in some cases are off by one order of magnitude, 
  while nonlinear  RF harvesting models from recent  prior art,  that do not take into account  limited harvesting sensitivity, 
  deviate from  reality  in the  low-input-power regime. 
 \item The proposed methodology can be applied to any type of RF energy harvesting system, provided that   system-level 
 datapoint pairs of the harvested output power and  the input power are provided. In that way, accurate SWIPT analysis can be facilitated.  
 \end{itemize}

The rest of the document is organized as follows.
Section~\ref{sec:Signal Model} introduces the channel model,
Section~\ref{sec:fund_EH_Baseline_methods} presents the  fundamentals of   far field RF energy  harvesting, explaining the inherent nonlinearity 
in the  real energy harvesting models. Section~\ref{sec:proposed_harv_model} presents the 
proposed approximation methodology, while Section~\ref{sec:energy_harv_applic} 
compares baseline, linear  harvesting models used in prior art with the nonlinear harvesting model,
under three performance metrics. Finally, work is concluded in Section~\ref{conclusion}.


\emph{Notation}: 
The set of natural and real numbers is denoted as $\mathds{N}$ and $\mathds{R}$, respectively.
For a natural number $N \in \mathds{N}$,  set $\{1,2, \ldots, N\}$ is denoted as $[N]\triangleq \{1,2, \ldots, N\}$.  
Random variables (RVs) are denoted with bold italic letters, e.g., $\pmb{a} $, while vectors
are denoted with underlined bold letters, e.g., $\mathbf{\underline{b}}$. Notation $\mathbf{\underline{b}}[j]$ stands for the $j$-th element of vector 
$\mathbf{\underline{b}}$.
Symbol $\odot$ stands for the  component-wise (Hadamard) product.
Notation  $\mathcal{CN}(0, \sigma^2)$ stands for the circularly-symmetric complex Gaussian distribution of variance $\sigma^2$.  
For a continuous  RV $\pmb{a}$, supported over an interval  set $\mathcal{X} $,
the corresponding PDF and  CDF is denoted  as $\mathsf{f}_{ \pmb{a}}(\cdot)$ and $\mathsf{F}_{ \pmb{a}}(x_0) = \int_{y\in \mathcal{X}: y\leq x_0} \mathsf{f}_{\pmb{a}}(y) \mathsf{d}y$,
respectively. The expectation and variance of $\mathsf{g}(\pmb{a})$ is denoted as  $\underset{   }{\mathbb{E}} [ \mathsf{g}(\pmb{a})]$ and $\mathsf{var}[ \mathsf{g}(\pmb{a})]
\triangleq \mathbb{E}[(\mathsf{g}(\pmb{a}) - \mathbb{E} [ \mathsf{g}(\pmb{a})])^2 ]$, respectively. 
The Dirac delta function is denoted as   
 $  \sDelta (\cdot)$.
The probability of event $\mathscr{S}$ is denoted as $\mathbb{P}(\mathscr{S})$  and ${\bf dom} \mathsf{g}$ denotes the domain of function $\mathsf{g}$. 

\vspace{-0.1 in}
\section{Wireless System Model}
\label{sec:Signal Model}

A source of RF signals offers wireless power to an information and far field RF energy harvesting (IEH) terminal. 
The source of   RF signals is assumed with a dedicated
power source, while the far field IEH terminal harvests RF energy 
from the incident signals on its antenna and could  
operate as information transmitter or receiver.

Narrowband transmissions are considered
over a quasi-static flat fading channel. For a single channel use, the downlink received signal 
at the output of the matched filter at the  IEH terminal is given by:  
\begin{equation}
\pmb{y}  =\sqrt{ P_{\rm T} \, T_{\rm s} \, \mathsf{L}(d)} \,\pmb{h} \,  \pmb{s} +   \pmb{w} ,
\end{equation}   
where $\pmb{s}$ is the transmitted symbol, with
$\mathbb{E} [\pmb{s}] = 0$ and  $\mathbb{E} \left[ |\pmb{s}|^2 \right]=1$, 
 $ P_{\rm T}$  is the average transmit power of the RF source, $T_{\rm s}$ 
 is the symbol duration, 
$\pmb{h}$ is the  complex baseband channel response, 
$ \mathsf{L}(d)$ is the path-gain (or inverse path-loss) coefficient at distance $d$,
and $\pmb{w} \sim \mathcal{CN}(0, \sigma_{\rm d}^2)$ is the 
additive white complex Gaussian noise at the IEH receiver.

A block fading model is considered, where the channel response
changes independently every coherence block of $T_{\rm c}$ seconds.
$\pmb{h}^{(n)}$ denotes the complex baseband channel response 
at the $n$-th coherence block. At each coherence block, 
the RF source transmits a packet whose duration
spans  $T_{\rm p}$ seconds, which in turn spans several symbols, with $T_{\rm p} \leq T_{\rm c}$.
The received RF input power (simply abbreviated as \emph{input power}) at the IEH terminal during the $n$-th coherence time block  is given by:
\begin{equation}
 \pmb{P}_{\rm R}^{(n)}    =  \mathbb{E}\! \left[ |\pmb{s}|^2 \right]  P_{\rm T}\,  \mathsf{L}(d)\,  \left|\pmb{h}^{(n)}\right|^2=  
\mathsf{P}(d)  \pmb{\gamma}^{(n)}  ,\label{eq:avg_rec_power}
\end{equation}
where $\mathsf{P}(d) \triangleq  P_{\rm T}\,  \mathsf{L}(d)$ and         $\pmb{\gamma}^{(n)} \triangleq   \left |\pmb{h}^{(n)} \right |^2$.
Note that $\pmb{P}_{\rm R}^{(n)}  $ is a function of  $\pmb{\gamma}^{(n)}$, i.e., $\pmb{P}_{\rm R}^{(n)} \equiv \pmb{P}_{\rm R}^{(n)}(\pmb{\gamma}^{(n)}) $.
Due to the definition of channel 
coherence time block, RVs $\left\{\pmb{h}^{(n)} \right\}$   are independent and identically 
distributed (IID) across different values of $n$. It is also assumed that  RVs $\pmb{\gamma}^{(n)} $ are drawn from a continuous
distribution, denoted as $\mathsf{f}_{\pmb{\gamma}^{(n)}}(\cdot)$, supported over the
non-negative reals, $\mathds{R}_+$.
Hence, the corresponding distribution of $ \pmb{P}_{\rm R}^{(n)}$
has a continuous density in $\mathds{R}_+$. 

 \begin{figure}[!t]
 \centering
         \includegraphics[width=0.75\columnwidth]{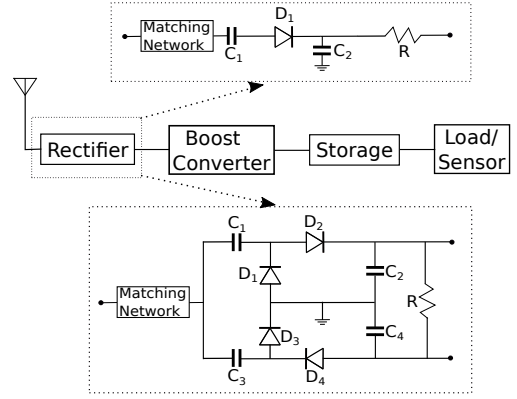}
  \caption{The architecture of  far field RF energy harvesters. Typical rectifier circuits with a single diode \cite{AsDaBle:16} (upwards)
  or multiple diodes \cite{OlChVol:11} (downwards) are also depicted, emphasizing  in the nonlinear relationship between   harvested and input  RF  power.}
 \label{fig:block_diagram_rectennas}
 \end{figure}
%
 
The presented results will be offered without having in mind a specific type of fading distribution. For the specific numerical results,  
Nakagami fading will be considered, since it can describe small-scale wireless fading under both line-of-sight (LoS) or
  non-line-of-sight (NLoS)  scenarios.  Under Nakagami distribution, the PDF of  $\pmb{\gamma}^{(n)} $ 
follows Gamma distribution  with shape  parameters $\left( \mathtt{m} , \frac{\Omega}{\mathtt{m}}\right)$,
given by:
\begin{equation}
 \mathsf{f}_{\pmb{\gamma}^{(n)} }(x)   =  \left( \frac{\mathtt{m}}{\Omega}\right)^{\mathtt{m}} \,
 \frac{x^{ \mathtt{m}- 1}}{\Gamma(\mathtt{m})} \, \mathsf{e}^{-\frac{\mathtt{m}}{\Omega} x}, ~~x \geq 0, 
\label{eq:gamma_distr}
\end{equation}
where  $\Gamma(x) = \int_{0}^{\infty}t^{x-1} \mathsf{e}^{-t} \mathsf{d}t$ is the Gamma function,
 while the Nakagami parameter $\mathtt{m}$ satisfies $\mathtt{m}  \geq \frac{1}{2}$. Parameter $\Omega$ satisfies 
 $\Omega = \mathbb{E}\!\left[\left |\pmb{h}^{(n)} \right |^2 \right] = \mathbb{E}\left[\pmb{\gamma}^{(n)}\right ]$.
For the special cases of $\mathtt{m} =1$ and  $\mathtt{m} =\infty$, Rayleigh and no-fading   is  obtained, respectively. 
For $\mathtt{m}= \frac{(\kappa + 1)^2}{2\kappa+ 1}$ 
 the distribution in Eq.~\eqref{eq:gamma_distr}
is approximated by a Rician distribution, with Rician parameter $\kappa $ \cite{Goldsmith:05}. The corresponding CDF of
RV $\pmb{\gamma}^{(n)}$ is given by:
\begin{equation}
 \mathsf{F}_{\pmb{\gamma}^{(n)} }(x)   = 1 - \int_x^{\infty} 
 \mathsf{f}_{\pmb{\gamma}^{(n)} }(y)  \mathsf{d} y = 1- 
\frac{\sGamma\!  \left(  \mathtt{m},  \frac{ \mathtt{m}}{ \Omega} x \right)}{\Gamma(\mathtt{m})} ,
  ~~x \geq 0, 
\label{eq:gamma_CDF}
\end{equation}
where   $\sGamma \!  \left(  \alpha ,  z \right) =  \int_{z}^\infty  t^{\alpha-1} \mathsf{e}^{-t} \mathsf{d}t   $ 
is the upper incomplete gamma function.
For exposition simplification, $\Omega= 1$ is assumed and thus, the input power, $\pmb{P}_{\rm R}^{(n)}$, in Eq.~\eqref{eq:avg_rec_power}  follows
Gamma distribution with shaping parameters $\left( \mathtt{m} , \frac{\mathsf{P}(d)}{\mathtt{m}}\right)$.

Finally, the following path-loss model is considered \cite{Goldsmith:05}: 
\begin{equation}
\mathsf{L}(d) = \left( \frac{\lambda}{d_0\, 4 \, \pi}\right)^2  \left( \frac{d_0}{d}\right)^{\nu} ,
\label{path_loss_model}
\end{equation}
with reference distance $d_0 =1$,  propagation wavelength $\lambda =0.3456$ and path-loss exponent (PLE) $\nu$.

\section{Fundamentals of Far Field RF Energy Harvesting}
\label{sec:fund_EH_Baseline_methods}

This section offers the fundamentals in RF energy harvesting, {filling a gap} largely overlooked in the recent wireless
communications theory prior art. The core of the far field RF energy harvesting circuit is the \emph{rectenna}, 
i.e., antenna and rectifier, that converts the incoming   RF signal to DC   under a nonlinear operation, commonly implemented with
one or more diodes. Increasing the number of diodes usually improves the harvesting efficiency, at the expense of reduced
harvesting sensitivity, explained below. 
Typical examples of rectifier circuits found in the literature are illustrated in Fig.~\ref{fig:block_diagram_rectennas}. 
A boost converter may be also incorporated after the rectifier, in order to amplify the required voltage and also offer 
maximum power point tracking (MPPT), exactly because the output of the rectifier is a nonlinear function of the   input power,
$\pmb{P}_{\rm R}^{(n)}$  \cite{KonkouMitBl:16}. It is apparent that accurate modeling of the nonlinearity in the harvester is of vital importance in joint
studies of the information and wireless power transfer  \cite{AsDaBle:16}, 
and that motivates this work. 

\subsection{Realistic Far Field RF Energy Harvesting Model}
\label{subsec:real_harv_model}

 \begin{figure*}[!t]
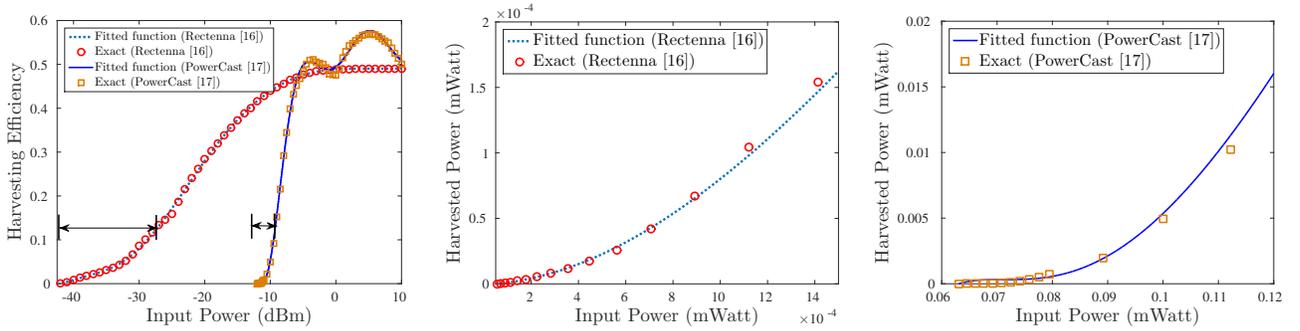

 \centering
         \includegraphics[width=0.605\columnwidth]{fig3.eps}
 \quad    \includegraphics[width=0.5905\columnwidth]{fig4.eps} 
 \quad     \includegraphics[width=0.622\columnwidth]{fig5.eps} 
  \caption{Left: The efficiency  of RF harvesting circuit as a function of input power in dBm
  for (a) the rectenna proposed  in \cite{AsDaBle:16}, depicted with circles and (b)
the PowerCast module \cite{PowerCast} (at $868$ Mhz), depicted with squares.
Center (Right): Harvested power vs. input power in mWatt for input power values depicted with arrows
in the left figure for the rectenna in \cite{AsDaBle:16} (module in \cite{PowerCast}).}
 \label{fig:efficiency_P_harv_vs_P_rec}
 \end{figure*}

The proposed ground-truth model for the  harvested power at the output of the RF harvesting 
circuit is  given by: 
\begin{equation}
\pmb{P}_{\rm har}^{(n)} \equiv \pmb{P}_{\rm har}^{(n)}\!\left( \pmb{P}_{\rm R}^{(n)} \right)     =  \mathsf{p}\! \left(  \pmb{P}_{\rm R}^{(n)}  \right), 
\label{eq:harvested_power_realistic}
\end{equation}
where
\begin{equation}
\mathsf{p}(x)
\triangleq 
\begin{cases}
0  , & x  \in [0,\mathtt{P}_{\rm in}^{\rm sen}], \\
\mathsf{e}\! \left( x  \right) \cdot x, & x \in [\mathtt{P}_{\rm in}^{\rm sen}, \mathtt{P}_{\rm in}^{\rm sat}] ,\\
 \mathsf{e}\! \left(  \mathtt{P}_{\rm in}^{\rm sat}\right) \cdot   \mathtt{P}_{\rm in}^{\rm sat} &x \in [\mathtt{P}_{\rm in}^{\rm sat}, \infty),
\end{cases}
\label{eq:harvested_power_realistic_function}
\end{equation}
where $x$ and $\mathsf{p}(x)$ take values in mWatt.
Function $ \mathsf{e}\! \left( \cdot \right)$ is   the harvesting efficiency as a function of
input  power, defined over  the interval  $
\mathcal{P}_{\rm in} \triangleq [\mathtt{P}_{\rm in}^{\rm sen}, \mathtt{P}_{\rm in}^{\rm sat}].
$
$\mathtt{P}_{\rm in}^{\rm sen}$ stands for
harvester's sensitivity;  for any input power value smaller
than sensitivity,  the harvested power is zero, i.e., $\mathsf{p}(x) = 0$
for $x \leq \mathtt{P}_{\rm in}^{\rm sen} $.
$\mathtt{P}_{\rm in}^{\rm sat}$ denotes the saturation power threshold of the harvester, 
after which the harvested power is constant.

Harvested power function  $ \mathsf{p}:\mathds{R}_+ \longrightarrow\mathds{R}_+ $
  is  assumed:
\begin{enumerate}
\item non-decreasing, i.e., $x<y \implies  \mathsf{p}(x) \leq  \mathsf{p}(y)$, and   
\item continuous,  i.e., $x \longrightarrow x_0 \implies  \mathsf{p}(x) \longrightarrow \mathsf{p}(x_0)$. 
\end{enumerate}  
Note that the assumptions above, even though mild, are in full accordance with  the
harvested power curves reported  in the RF energy harvesting circuits' prior art, e.g., \cite{VaDu:14, PopFaErCoZa:13, AsDaBle:16,PowerCast}.

Determining an explicit formula for   $\mathsf{p}(\cdot)$ in~\eqref{eq:harvested_power_realistic_function}, for a given rectifier circuit,
is crucial task and  requires first to specify the harvesting efficiency function  $ \mathsf{e}\! \left( \cdot \right)$  over the input power interval $\mathcal{P}_{\rm in}$.
Inline  with the prior art \cite{BoNgZlSc:15, XuOzAyMcKVi:17, BoNgZlSc:17_a,  BoNgZlSc:17_b, BoNgZlSc:17_c, BoNgZlSc:17_d}, 
for a given  rectifier circuit,  some  measured harvesting efficiency 
data points  are assumed available,   corresponding to some input power values (between sensitivity and saturation).
Assuming specific parametrization  for $ \mathsf{e}\! \left( \cdot \right)$ (e.g., polynomial, sigmoid functions), 
the measured harvesting efficiency data  can be harnessed to designate the best shape for    function  $ \mathsf{e}\! \left( \cdot \right)$
through parameter fitting.

In this work, the ground-truth harvesting efficiency function 
is modeled as a high-order polynomial in the dBm scale:
  \begin{align}
 \mathsf{\mathsf{e}}(x) = w_0 + \sum_{i=1}^{W}  w_i  (10 \, \mathsf{log}_{10}(x) )^i, ~x\in \mathcal{P}_{\rm in}.
 \label{eq:harvesting_efficiency_function}
\end{align}
Function in~\eqref{eq:harvesting_efficiency_function} is 
parametrized by $W+1$ real numbers -- the coefficients of the polynomial -- where  $W$ is the degree of the polynomial.
The  best values  for the coefficients $\{w_i\}_{i=0}^W$ can be found from the rectenna's measured harvesting efficiency data,   
exploiting standard convex optimization fitting techniques from \cite[Chapter~6]{BoydVand:04}.
The   optimized fitted function    $\mathsf{\mathsf{e}}(x)$ is non-negative and continuous over $\mathcal{P}_{\rm in}$ and obtains the value zero 
 for $x = \mathtt{P}_{\rm in}^{\rm sen}$.
The main benefit  of the proposed harvesting efficiency parametrization  in~\eqref{eq:harvesting_efficiency_function}
is the utilization of dBm scale, that  offers higher granularity over the very small input power values. 
  It is emphasized that Eq.~\eqref{eq:harvested_power_realistic_function}
will be only used for evaluation of the simplified piecewise linear approximation (proposed in the next section), 
based on datapoint pairs of harvested power  and corresponding input power. 

 Two rectenna  models from the RF harvesting circuit design prior art \cite{AsDaBle:16} and \cite{PowerCast} are evaluated.
 The first one is an ultra sensitive rectenna from the microwave theory prior art, while the latter is the PowerCast module.
 The range of the input power values for the rectenna models \cite{AsDaBle:16}  and \cite{PowerCast} 
 were $\mathcal{P}_{\rm in} = [10^{-4.25}, 10^{1.6}]$ mWatt and $\mathcal{P}_{\rm in} = [10^{-1.2}, 10]$  mWatt, respectively.
The number of the provided measured data   for the rectenna in \cite{AsDaBle:16} (PowerCast module \cite{PowerCast}) were
$118$  and ($53 $) points.
Fig.~\ref{fig:efficiency_P_harv_vs_P_rec}-Left illustrates the  harvesting efficiency as a function of input 
power in dBm of the two studied rectenna models.
Fig.~\ref{fig:efficiency_P_harv_vs_P_rec}-Center (Right)
illustrates the harvested power as a function of the input  power in mWatt, for the rectenna in \cite{AsDaBle:16} (harvester in
\cite{PowerCast}) and the input power range marked with arrows in Fig.~\ref{fig:efficiency_P_harv_vs_P_rec}-Left; 
it becomes clear that the  harvested  power is a nonlinear function of the input power. 
For the  rectenna   models  in \cite{AsDaBle:16}  
and     \cite{PowerCast},
the degrees of the fitted polynomials  for the  function  $\mathsf{\mathsf{e}}(x)$ are $W = 10$ and $W=12$, respectively (depicted in Fig.~\ref{fig:efficiency_P_harv_vs_P_rec}
with dotted and solid curves, respectively).\footnote{  The fitted polynomials  (in dBm scale) for the  two studied rectenna models 
are  provided online in \url{http://users.isc.tuc.gr/~palevizos/palevizos_links.html}.}


\subsection{Impact of Harvester's Sensitivity in RF Energy Harvesting}
\label{subsec:impact_sensitivity_harv_model}

 The harvester's sensitivity is a very important parameter playing vital role on the performance of the 
 rectenna. The sensitivity is the power threshold beyond which the rectifier is able to harvest RF energy and 
 depends on diode's turn-on (or threshold) voltage $\mathtt{V}_{\rm T}$, i.e., the voltage above which the diode is
said to be forward-biased \cite{VaDu:14}. As the turn-on
threshold voltage is decreased, the energy conversion
efficiency at a given power increases, i.e., the rectifier becomes more sensitive.

Unfortunately, prior art neglects the impact of harvester's sensitivity.
To this end, we define an important RF harvesting  metric,  given by
\begin{equation}
 \mathbb{P}(\pmb{P}_{\rm R} \leq \mathtt{P}_{\rm in}^{\rm sen}), 
 \label{eq:sens_outage_event}
\end{equation}
which is the probability that the input power (depending on the wireless channel) is below the  harvester's sensitivity
(depending on the harvester). Note that the probability event of~\eqref{eq:sens_outage_event}
is the fraction of  time  the rectenna cannot harvest RF energy
due to   inadequate  incident  input RF power.

 \begin{figure}[!t]
 \centering
         \includegraphics[width=0.7\columnwidth]{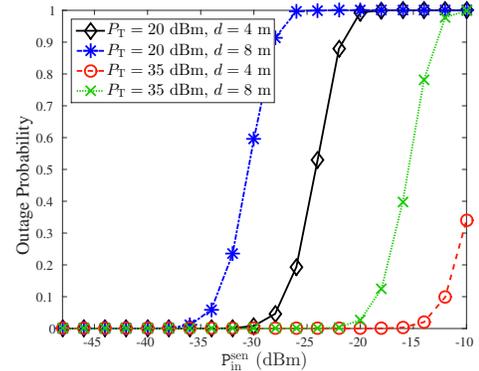}
  \caption{Probability of outage event as a function of harvester's sensitivity.}  
 \label{fig:impact_sensitivity}
 \end{figure}
 
 Fig.~\ref{fig:impact_sensitivity} examines the probability of outage in Eq.~\eqref{eq:sens_outage_event} as a function of  
 harvester's sensitivity, $\mathtt{P}_{\rm in}^{\rm sen}$. 
 The path-loss model of Eq.~\eqref{path_loss_model} is employed with  $\nu = 2.1$
 and Nakagami parameter  $\mathtt{m}=5$. It can be clearly deduced that the smaller
  the harvester's sensitivity is, the larger the outage probability in~\eqref{eq:sens_outage_event} becomes.
 Thus, for the less sensitive PowerCast module  \cite{PowerCast},
 the probability of outage due to limited input power is almost $1$ for transmission power $P_{ \rm T} = 20$ dBm
 and transmitter-receiver distance $d$ more than $4$ meters, while for   $P_{ \rm T} = 35$ dBm and $d = 4$ meters
 the outage event   becomes $10\%$. For the sensitive
 rectenna in  \cite{AsDaBle:16}  the outage event becomes almost $0$ for all studied scenarios for the  parameters   $P_{ \rm T}$  and $d$.
 We conclude that the less sensitive the rectenna is, the major    the impact
 of   harvester's sensitivity becomes on the accuracy of the studied RF harvesting model, especially   in the    low-input-power   regime.

\subsection{Prior Art (Linear) RF Energy Harvesting Models}
\label{subsec:baseline_harv_model}


Three baseline models are considered for comparison:

 \subsubsection{Linear  (L)  Energy Harvesting Model}
\label{subsec:linear_harv_model} 

The first baseline  model is the linear (L) model adopted by a gamut
of information and wireless energy transfer prior art; for that model, the harvested power (as function of $\pmb{P}_{\rm R}^{(n)}$) is expressed as follows:
\begin{equation}
    {\mathsf{p}}_{\rm L}\! \left(  x \right) = \eta_{\rm L}  \cdot x , ~   x  \in \mathds{R}_+,
\label{eq:harvested_power_linear}
\end{equation}
with  constant $\eta_{\rm L} \in [0,1)$. The functional form of the harvested power in \eqref{eq:harvested_power_linear}
is depicted in Fig.~\ref{fig:problem} with solid curve.
This model ignores the following:
(i) the dependence of RF harvesting efficiency on input  power,
(ii) the harvester cannot operate below the sensitivity threshold,  and 
(iii) the  harvested power saturates  when the input power level is above a power threshold.
 
 \subsubsection{Constant-Linear  (CL)  Energy Harvesting Model}
\label{subsec:constant_linear_harv_model} 
The harvested power  is expressed as follows:
\begin{equation}
   {\mathsf{p}}_{\rm CL}\! \left(  x  \right) \triangleq 
\begin{cases}
0  , &  x\in [0,\mathtt{P}_{\rm in}^{\rm sen}], \\
 \eta_{\rm CL} \cdot  (x- \mathtt{P}_{\rm in}^{\rm sen})  , &  x \in [\mathtt{P}_{\rm in}^{\rm sen}, \infty) ,
\end{cases}
\label{eq:harvested_power_constant_linear}
\end{equation}
with   constant $\eta_{\rm CL} \in [0,1)$. The CL  harvested power curve is depicted
with dash-dotted  line  in Fig.~\ref{fig:problem}.
This model takes into account the fact that the RF harvester is not able to operate
  below  sensitivity threshold $\mathtt{P}_{\rm in}^{\rm sen}$.
On the contrary, the CL model  ignores that 
 RF harvesting efficiency is a non-constant  function of input power 
and  that the  harvested power saturates  when  the input power  is above   $\mathtt{P}_{\rm in}^{\rm sat}$.

 \subsubsection{Constant-Linear-Constant  (CLC) Energy Harvesting Model}
\label{subsec:constant_linear_constant_harv_model} 

The harvested power is expressed as a function of  input power $\pmb{P}_{\rm R}^{(n)}$, 
through the following expression:
\begin{equation}
 {\mathsf{p}}_{\rm CLC}\! \left(  x  \right) \triangleq 
\begin{cases}
0  , & x \in [0,\mathtt{P}_{\rm in}^{\rm sen}], \\
 \eta_{\rm CLC}\cdot (x-\mathtt{P}_{\rm in}^{\rm sen})  , & x \in [\mathtt{P}_{\rm in}^{\rm sen}, \mathtt{P}_{\rm in}^{\rm sat}] ,\\
 \eta_{\rm CLC}\cdot (\mathtt{P}_{\rm in}^{\rm sat}  - \mathtt{P}_{\rm in}^{\rm sen}),&  x  \in [\mathtt{P}_{\rm in}^{\rm sat}, \infty),
\end{cases}
\label{eq:harvested_power_constant_linear_constant}
\end{equation}
where constant $\eta_{\rm CLC} \in [0,1)$.  
The CLC model is depicted in Fig.~\ref{fig:problem} with a  dotted  curve.
This last model ignores the dependence of harvesting efficiency on input power. 
In our simulation scenarios, parameters 
$\eta_{\rm L}$, $\eta_{\rm CL}$, and $\eta_{\rm CLC}$ have 
been chosen empirically to minimize their performance mismatch compared to the real RF
harvesting model in Eq.~\eqref{eq:harvested_power_realistic}.

\section{Statistics of Harvested Power}
\label{sec:proposed_harv_model}

 \begin{figure}[!t]
 \centering
         \includegraphics[width=0.8\columnwidth]{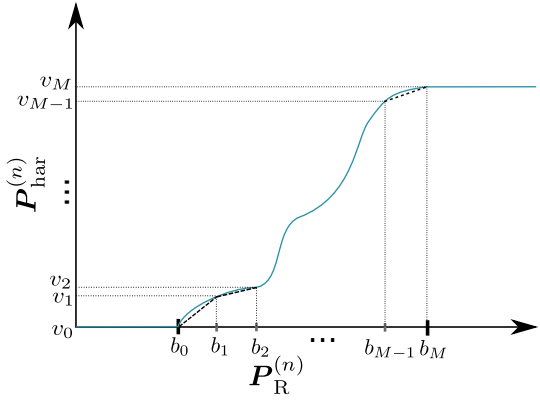}
  \caption{A graphical illustration of the proposed piecewise linear approximation for an RF energy harvesting model, 
  adhering to the mild assumptions of Section~\ref{subsec:real_harv_model}.}
 \label{fig:P_harv_vs_P_rec_proposed_model}
 \end{figure}


Consider  the harvesting model in   Eq.~\eqref{eq:harvested_power_realistic}  where the function $\mathsf{p}(\cdot)$
satisfies the assumptions in Section~\ref{subsec:real_harv_model}.
The proposed methodology uses a piecewise linear approximation of
  $\mathsf{p}(\cdot)$ over interval  $\mathcal{P}_{\rm in} $  using $M+1$ points. 

Since the harvested power $\pmb{P}_{\rm har}^{(n)}$    in Eq.~\eqref{eq:harvested_power_realistic} changes over the range
of input power values    $\mathcal{P}_{\rm in} $,
a set of  support points $\{b_m\}_{m=0}^{M}$ is defined, with
$b_0 = \mathtt{P}_{\rm in}^{\rm sen}$, $b_{m-1} < b_{m}$, for $m \in [M]$,
and $b_M =  \mathtt{P}_{\rm in}^{\rm sat}$. 
The corresponding set of image points    $\{v_m \}_{m=0}^{M} \triangleq \{\mathsf{p}(b_m)\}_{m=0}^{M}$
satisfy $ v_{m-1} =    \mathsf{p}(b_{m-1}) \leq \mathsf{p}(b_m) = v_m$,  $m=1,2,\ldots, M$,
with  $v_0  = 0 $
 and  $v_M = \mathsf{p}( \mathtt{P}_{\rm in}^{\rm sat})$. 
Without loss of generality, $0 = v_0< v_1 < v_2 < \ldots <v_{M-1} < v_M  = \mathsf{p}( \mathtt{P}_{\rm in}^{\rm sat})$ is assumed.
The methodology is graphically illustrated in Fig.~\ref{fig:P_harv_vs_P_rec_proposed_model}.

Given the   $M+1$ points  $\{b_m\}_{m=0}^{M}$ and
$\{v_m\}_{m=0}^{M}$, slopes $l_m \triangleq \frac{v_{m} - v_{m-1}}{b_m - b_{m-1}}$, $m \in [M]$ are defined.
The utilized methodology  approximates   $\pmb{P}_{\rm har}^{(n)} $ in   Eq.~\eqref{eq:harvested_power_realistic}
through the following piecewise linear function:
\begin{equation}
\widetilde{\pmb {P}}_{\rm har}^{(n)} \equiv \widetilde{\pmb {P}}_{\rm har}^{(n)}\!\left( \pmb{P}_{\rm R}^{(n)}\right) 
=   \widetilde{\mathsf{p}}\!\left( \pmb{P}_{\rm R}^{(n)}\right)  
\label{eq:P_harv_approx}
\end{equation}
with
\begin{equation}
 \widetilde{\mathsf{p}}(x)
\triangleq  \begin{cases}
0 &  x   \in [0,  b_0], \\
l_m ( x -  b_{m-1}) + v_{m-1},  &  x \in ( b_{m-1}, b_m]   ,   m\in [M] ,\\
v_M , &  x \in  [ b_M, \infty).
\end{cases}
\label{eq:P_harv_approx_function}
\end{equation}

The computational complexity to evaluate the   function in~\eqref{eq:P_harv_approx_function}
is $\mathcal{O}(M)$.  On the other hand, $\mathcal{O}(1)$ computational cost is required 
to evaluate the baseline models   in 
Eqs.~\eqref{eq:harvested_power_linear}--\eqref{eq:harvested_power_constant_linear_constant},
the proposed harvested power function in Eq.~\eqref{eq:harvested_power_realistic_function}, as well as
the harvested power functions from the nonlinear RF harvesting   prior art \cite{BoNgZlSc:15, XuOzAyMcKVi:17, BoNgZlSc:17_a,  BoNgZlSc:17_b, BoNgZlSc:17_c, BoNgZlSc:17_d}. 
However, the focus in this work is to assess
important RF harvesting performance evaluation metrics in nonlinear RF harvesting,
and thus, the computational cost is not a critical issue. One important benefit of the  piecewise linear 
approximation in~\eqref{eq:P_harv_approx}    based on measured input-output datapoints,  is its  flexibility 
to interpolate directly the harvested power values,   
without having the exact functional form of $\mathsf{p}(\cdot)$.  Thus,
one can directly assess important RF harvesting evaluation metrics without 
assuming a specific functional form for the harvested power function.

\subsection{Statistics of  $\widetilde{\pmb {P}}_{\rm har}^{(n)}$ and Approximation Error}
\label{sec:statistical_descr_P_harv}

This section offers the PDF and CDF of $\widetilde{\pmb {P}}_{\rm har}^{(n)}$.
First, the following is defined:
\begin{equation}
\xi_m \triangleq  \mathsf{F}_{ \pmb {P}_{\rm R}^{(n)}}(b_m), ~m=0,1,\ldots, M,
\label{eq:CDF_points}
\end{equation}
where $ \mathsf{F}_{ \pmb {P}_{\rm R}^{(n)}}(\cdot)$ is the CDF of $\pmb {P}_{\rm R}^{(n)}$.
From Eq.~\eqref{eq:P_harv_approx} it can be remarked that
$\widetilde{\pmb {P}}_{\rm har}^{(n)}  = 0$ with probability 
\begin{align}
& \mathbb{P}\!\left ( \pmb {P}_{\rm R}^{(n)} \leq  b_0 \right)
= \int_{0}^{b_0} \mathsf{f}_{\pmb {P}_{\rm R}^{(n)}}(x) \mathsf{d} x = \mathsf{F}_{\pmb {P}_{\rm R}^{(n)}}(b_0) = \xi_0   \nonumber \\
\implies &  \mathsf{f}_{\widetilde{\pmb {P}}_{\rm har}^{(n)}}(x) =  \xi_0 \, \sDelta(x) , ~ x = 0.
\label{eq:f_P_h_0}
\end{align}
For any $m \in [M-1]$, when $\pmb {P}_{\rm R}^{(n)} \in ( b_{m-1}, b_m] $,
$\widetilde{\pmb {P}}_{\rm har}^{(n)} \in (v_{m-1}, v_m] $ holds. Thus,
using the formula for  linear transformations   in \cite{pappoulis:02} 
the following is obtained  for any $m \in [M-1]$:   
\begin{align}
\mathsf{f}_{\widetilde{\pmb {P}}_{\rm har}^{(n)}}(x) = \frac{1}{l_m} \, \mathsf{f}_{\pmb {P}_{\rm R}^{(n)} }\!\! \left( 
\frac{x  -  v_{m-1} + l_m b_{m-1}}{l_m} \right),
\label{eq:f_P_h_v_m}
\end{align}
for $x  \in (v_{m-1}, v_m]$.
Note that the last interval $\pmb{P}_{\rm R}^{(n)} \in ( b_{M-1}, b_M] $ requires special attention 
 due to the fact that  the inverse of  function  $ \widetilde{\mathsf{p}}(\cdot)$  does not exist  at point  $v_M$.
Restricting   $\pmb{P}_{\rm R}^{(n)} \in ( b_{M-1}, b_M) $, the following holds:  
\begin{align}
\mathsf{f}_{\widetilde{\pmb {P}}_{\rm har}^{(n)}}(x) = \frac{1}{l_M} \, \mathsf{f}_{\pmb {P}_{\rm R}^{(n)} }\!\! \left( 
\frac{x  -  v_{M-1} + l_M b_{M-1}}{l_M} \right),  
\label{eq:f_P_h_v_M_minus1}
\end{align}
for $x  \in (v_{M-1}, v_M)$.
Finally, in view of~\eqref{eq:P_harv_approx}, $\widetilde{\pmb {P}}_{\rm har}^{(n)} = v_M$
with probability given by: 
\begin{align}
& \mathbb{P}\!\left ( \pmb{P}_{\rm R}^{(n)} \geq   b_M \right) = 1 - \lim_{x \rightarrow  b_M}\mathsf{F}_{\pmb{P}_{\rm R}^{(n)}}(x) 
\overset{(a)}{=} 1 - \xi_M \nonumber \\
\implies &  \mathsf{f}_{\widetilde{\pmb {P}}_{\rm har}^{(n)}}(x) =  (1- \xi_M) \, \sDelta(x-v_M) , ~ x = v_M,
\label{eq:f_P_h_v_M}
\end{align}
where $(a)$ stems from the continuity of  $\mathsf{F}_{\pmb{P}_{\rm R}^{(n)}}(\cdot)$ 
as an integral function of a continuous PDF   \cite{Fol:99}, 
as well as the definition of   $ \xi_M$ in~\eqref{eq:CDF_points}. 
Thus, the following proposition summarizes the results related
to the probabilistic description of  $\widetilde{\pmb {P}}_{\rm har}^{(n)}$.

\begin{proposition} \normalfont
For   a given  distribution   of  the fading power $\pmb{\gamma}^{(n)}$, 
supported over $\mathds{R}_+$, in view of Eq.~\eqref{eq:avg_rec_power}, 
the corresponding distribution of 
  the  input power, $\pmb{P}_{\rm R}^{(n)}$,  is  $\mathsf{f}_{\pmb{P}_{\rm R}^{(n)}}(x) = 
    \frac{1}{\mathsf{P}(d)}\mathsf{f}_{\pmb{\gamma}^{(n)}}\!\!\left(\frac{x}{\mathsf{P}(d)} \right) $.
Hence, the  proposed  approximation  in Eq.~\eqref{eq:P_harv_approx}
has PDF:
\begin{align}
&\mathsf{f}_{\widetilde{\pmb {P}}_{\rm har}^{(n)}}(x)   \nonumber \\
=&~ 
\begin{cases}
  \xi_0 \, \sDelta(x) , & x = v_0 = 0 , \\
  \frac{1}{l_m} \, \mathsf{f}_{\pmb{P}_{\rm R}^{(n)} }\!\! \left( 
\frac{x  -  v_{m-1} + l_m b_{m-1}}{l_m} \right),  & x  \in (v_{m-1}, v_m]\backslash \{ v_M\},m \in [M], \\
 (1- \xi_M) \, \sDelta(x-v_M) , & x = v_M, \\
0, & x \in \mathds{R}  \backslash [0, v_M],
\end{cases}
\label{eq:PDF_P_harv}
\end{align} 
where  $m \in [M]$.
The corresponding CDF of  $\widetilde{\pmb {P}}_{\rm har}^{(n)}$  is
given by:
\begin{align}
&\mathsf{F}_{\widetilde{\pmb {P}}_{\rm har}^{(n)}}(x)  \nonumber \\  
=&~
\begin{cases}
  0  & x  <0, \\
    \mathsf{F}_{\pmb{P}_{\rm R}^{(n)} }\!\! \left( 
\frac{x  -  v_{m-1} + l_m b_{m-1}}{l_m} \right),  & x  \in  [v_{m-1}, v_m] \backslash \{ v_M\} ,m \in [M], \\
1 , & x \geq v_M.
\end{cases}
\label{eq:CDF_P_harv}
\end{align} 
\begin{proof}
The proof of Eq.~\eqref{eq:PDF_P_harv} is immediate from Eqs.~\eqref{eq:f_P_h_0}--\eqref{eq:f_P_h_v_M}.
The proof  of Eq.~\eqref{eq:CDF_P_harv} is given  in Appendix~\ref{sec:proof_prop_CDF_Pharv}.
\end{proof}
\label{prop:PDF_CDF_Pharv}
\end{proposition}

It is shown immediately below that the proposed approximation in  Eq.~\eqref{eq:P_harv_approx_function}   offers    approximation error that decays 
quadratically with the number of utilized points, even for a  uniform  choice of points
$\{b_m\}$, i.e.,  $b_m = b_{m-1}+\delta_M$, $m \in [M]$,
with $\delta_M \triangleq \frac{ \mathtt{P}_{\rm in}^{\rm sat} - \mathtt{P}_{\rm in}^{\rm sen}}{M} $.

\begin{proposition}[Approximation Error with Uniform Point Selection] \normalfont
Suppose that    we choose $b_m = b_{m-1}+\delta_M$, $m \in [M]$,
with $\delta_M$ defined as  above.
 If  the function $\mathsf{p}(\cdot)$ is in addition continuously differentiable,
 then  $ \widetilde{\mathsf{p}}(\cdot)$  in~\eqref{eq:P_harv_approx_function},  
restricted over $\mathcal{P}_{\rm in} $, approximates    $\mathsf{p}(\cdot)$, over  $\mathcal{P}_{\rm in} $, 
with an absolute error that is bounded  as follows:
\begin{equation}
\int_{\mathcal{P}_{\rm in} }  \left|  \mathsf{p}(x) -
\widetilde{\mathsf{p}}(x) \right|  \mathsf{d}x  \leq  
  \frac{  \mathtt{C}_{\mathsf{p}} \, (\mathtt{P}_{\rm in}^{\rm sat} - \mathtt{P}_{\rm in}^{\rm sen})^3}{8 \, M^2},
\label{eq:approximation_error_proposition}
\end{equation}
where  $\mathtt{C}_{\mathsf{p}} =
  \max_{x \in \mathcal{P}_{\rm in}}\left | \mathsf{p}''(x) \right|$ is a constant
independent of $M$. 
\begin{proof}
The proof is provided in Appendix~\ref{sec:proof_prop_approx_err}.
\end{proof}
\label{prop:approximation_error_poly}
\end{proposition}

Thus, at most
$\mathcal{O}\!\left(\sqrt{\frac{1}{\epsilon}}\right)$ number of support points is required to approximate the function $\mathsf{p}(\cdot)$ with accuracy  at least $\epsilon$.


\section{Evaluation}
\label{sec:energy_harv_applic}

\subsection{Baseline Comparison: Average Harvested Energy}
\label{subsec:adopted_harv_model}
%
For baseline comparison,  the expected harvested energy is considered. 
$\pmb{U}_N \triangleq \sum_{n=1}^N \pmb{P}_{\rm har}^{(n)}$ denotes the accumulated harvested power
up to coherence block $N$, which in turn offers the expected harvested energy over $N$ coherence periods:
\begin{equation}
\mathbb{E} \! \left[ T_{\rm p} \,
\pmb{U}_N
 \right] =  T_{\rm p} \,  \mathbb{E} \! \left[
\sum_{n=1}^N \pmb{P}_{\rm har}^{(n)}
 \right] =  N \,  T_{\rm p} \, \mathbb{E}  \! \left[ \pmb{P}_{\rm har}^{(n)}  \right],
\label{eq:expected_harvested_energy_N}
\end{equation} 
for some $n\in [N]$. The last equality stems from the fact that 
$ \{\pmb{P}_{\rm har}^{(n)}\}_{n \in [N]} $ are identically distributed, 
since $\{\pmb{\gamma}^{(n)}\}_{n \in [N]} $ are also identically distributed.
Let us denote $ \overline{\mathtt{ P}}_{\rm L}$, 
$ \overline{\mathtt{ P}}_{\rm CL}$, $ \overline{\mathtt{ P}}_{\rm CLC}$,  and
$ \overline{\widetilde{\mathtt{ P}}}_{}$
the expected  harvested power  over a single coherence block of  the following models, respectively:  linear
 in Eq.~\eqref{eq:harvested_power_linear},    constant-linear 
 in Eq.~\eqref{eq:harvested_power_constant_linear},
   constant-linear-constant  in Eq.~\eqref{eq:harvested_power_constant_linear_constant}, 
and  proposed  in Eq.~\eqref{eq:P_harv_approx}.

\begin{figure*}[!t]
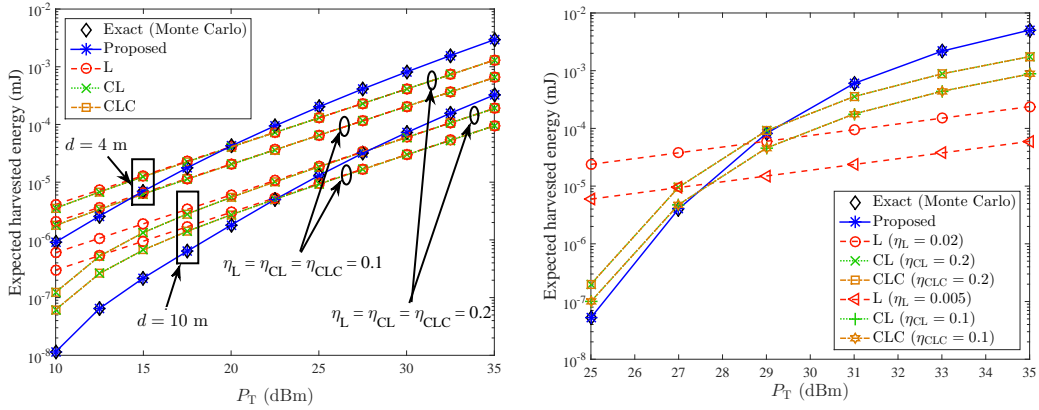

 \centering
         \includegraphics[width=0.75\columnwidth]{fig8.eps}  \quad
   \includegraphics[width=0.75\columnwidth]{fig9.eps} 
  \caption{Left (Right): Expected harvested energy per coherence block ($N=1$) vs. transmission power $P_{\rm T}$ for the rectenna proposed in  \cite{AsDaBle:16}
(harvesting module in \cite{PowerCast}).}
 \label{fig:EhE_vs_Pt}
 \end{figure*}

Under Nakagami fading, the average harvested power for the baseline linear models is
given by:
\begin{align} 
 \overline{\mathtt{ P}}_{\rm L} & =  \mathbb{E}[{\mathsf{p}}_{\rm L}(\pmb{P}_{\rm R})] =  \eta_{\rm L}\, \mathsf{P}(d) \\
 \overline{\mathtt{ P}}_{\rm CL} & =  \mathbb{E}[{\mathsf{p}}_{\rm CL}(\pmb{P}_{\rm R})] =
\int_{0 }^{\infty } \! {\mathsf{p}}_{\rm CL}(x) \,
  \mathsf{f}_{\pmb{P}^{(n)}_{\rm R} }(x) \mathsf{d}x \nonumber \\
& =   \eta_{\rm C L}  \left(    \frac{\mathsf{P}(d) \, \sGamma\!  \left(  
\mathtt{m}+1,  \frac{ \mathtt{m}}{ \mathsf{P}(d) } \mathtt{P}_{\rm in}^{\rm sen}  \right)}{\Gamma(\mathtt{m}+1)} -
\frac{  \mathtt{P}_{\rm in}^{\rm sen}   \,  \sGamma\!  \left(  \mathtt{m},
  \frac{ \mathtt{m}}{ \mathsf{P}(d)}\mathtt{P}_{\rm in}^{\rm sen}   \right)}{\Gamma(\mathtt{m})}
  \right)\\
 \overline{\mathtt{ P}}_{\rm CLC} & =  \mathbb{E}[{\mathsf{p}}_{\rm CLC}(\pmb{P}_{\rm R})] =  
\int_{0 }^{\infty } {\mathsf{p}}_{\rm CLC}(x) \,
  \mathsf{f}_{\pmb{P}^{(n)}_{\rm R} }(x) \mathsf{d}x  \nonumber \\
&=   \left( \frac{ \mathsf{P}(d)     \left(
 \sGamma\!  \left(  \mathtt{m}+1,  \frac{ \mathtt{m}}{ \mathsf{P}(d) } \mathtt{P}_{\rm in}^{\rm sen}  \right ) -
  \sGamma\!  \left(  \mathtt{m} +1,  \frac{ \mathtt{m}}{ \mathsf{P}(d) } \mathtt{P}_{\rm in}^{\rm sat}   \right)  \right)  }
{\Gamma(\mathtt{m}+1 )}  \right. \nonumber \\
 &  + \left. \! \frac{   \mathtt{P}_{\rm in}^{\rm sat}   \,  \sGamma\!  \left(  \mathtt{m},
  \frac{ \mathtt{m}}{ \mathsf{P}(d)}\mathtt{P}_{\rm in}^{\rm sat}   \right)}{\Gamma(\mathtt{m})} - \frac{  \mathtt{P}_{\rm in}^{\rm sen}   \,  \sGamma\!  \left(  \mathtt{m},
  \frac{ \mathtt{m}}{ \mathsf{P}(d)}\mathtt{P}_{\rm in}^{\rm sen}   \right)}{\Gamma(\mathtt{m})} \right)  \eta_{\rm C L C} ,
\end{align}
where   the expressions above rely on   $\Gamma(\mathtt{m}+1 )  =  \mathtt{m} \cdot \Gamma(\mathtt{m}) $, as well as on the following formula ($i \in \mathds{N} \cup \{0\}$) 
\cite[Eq.~(3.381.9)]{gradshteyn:07}:
\begin{align}
 \int_a^{b}\!\!  x^i  \,  \mathsf{f}_{\pmb{P}^{(n)}_{\rm R} }(x) \mathsf{d}x = 
\left(\frac{ \mathsf{P}(d)}{\mathtt{m}} \right)^{i} \, 
\frac{  
 \sGamma\!  \left(  \mathtt{m}+i,  \frac{ \mathtt{m}}{ \mathsf{P}(d) } a \right ) -
  \sGamma\!  \left(  \mathtt{m} +i,  \frac{ \mathtt{m}}{ \mathsf{P}(d) } b  \right)   }
{ \Gamma(\mathtt{m} )}.
\label{eq:aux_equation}
\end{align}

 \begin{figure*}[!t]
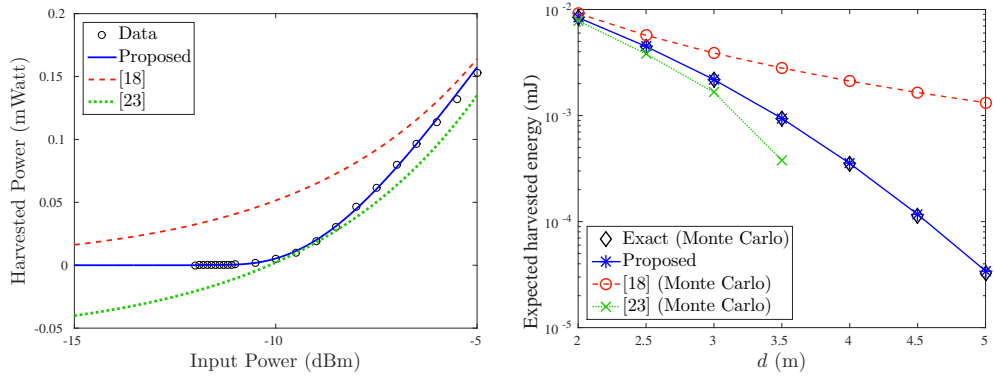

 \centering
          \includegraphics[width=0.71\columnwidth]{fig10.eps}   \quad
         \includegraphics[width=0.71\columnwidth]{fig11.eps}
  \caption{Left: Harvested power vs. input power for the proposed and the other nonlinear 
  RF harvesting models from the prior art using the harvesting module in \cite{PowerCast}. Right: Expected 
  harvested energy per coherence block ($N=1$) vs.  distance $d$.}
 \label{fig:EhE_vs_dist}
 \end{figure*}

For the {proposed} piecewise linear approximation, the expected harvested power over a single coherence period is given by:
 \begin{align}
\overline{\widetilde{\mathtt{ P}}}  & = \mathbb{E}[\widetilde{\mathsf{p}}(\pmb{P}_{\rm R})] \nonumber \\
&= \sum_{j=1}^M \left( \frac{ l_j \, \mathsf{P}(d)     \left(
 \sGamma\!  \left(  \mathtt{m}+1,  \frac{ \mathtt{m}\,  b_{j-1} }{ \mathsf{P}(d) }  \right ) -
  \sGamma\!  \left(  \mathtt{m} +1 ,  \frac{ \mathtt{m} \,  b_{j}}{ \mathsf{P}(d) }   \right)  \right)  }{\Gamma(\mathtt{m}+1 )} 
 \right. \nonumber \\ 
 &  \left. +  \frac{ (v_{j-1} -  l_j  b_{j-1} )
 \left(  \sGamma\!  \left(  \mathtt{m} ,  \frac{ \mathtt{m}\,  b_{j-1} }{ \mathsf{P}(d) }  \right ) -
  \sGamma\!  \left(  \mathtt{m},  \frac{ \mathtt{m} \,  b_{j}}{ \mathsf{P}(d) }   \right)  \right) }{\Gamma(\mathtt{m} )} \right)    \nonumber\\
  \label{eq:expected_harvested_power}  
& ~+ \frac{ v_M    \,  \sGamma\!  \left(  \mathtt{m},
  \frac{ \mathtt{m}}{ \mathsf{P}(d)}\mathtt{P}_{\rm in}^{\rm sat}   \right)}{\Gamma(\mathtt{m})},
\end{align}
where  Eq.~\eqref{eq:aux_equation} is exploited to 
obtain the  final simplified   expression.

%

\subsubsection{Numerical Results}

The expected harvested energy in Eq.~\eqref{eq:expected_harvested_energy_N} is found
for the actual energy harvesting model in  Eq.~\eqref{eq:harvested_power_realistic} (obtained through Monte Carlo experiments), 
for the three linear baseline models, and the proposed piecewise linear   energy harvesting model.

 Fig.~\ref{fig:EhE_vs_Pt} examines the impact of transmit power $P_{\rm T}$ on the average  harvested energy 
over $N=1$ coherence period using  $T_{\rm p} = 50$ msec. In Fig.~\ref{fig:EhE_vs_Pt}-Left,   $\nu=2.1$ and $\mathtt{m}=5$ 
are set, for the rectenna in \cite{AsDaBle:16}. It can be observed that the expected harvested energy performance of the proposed 
approximation in~\eqref{eq:P_harv_approx} with   $M+1 = 586$  points
is the same with the performance of the actual harvesting model for all studied distance scenarios of $d=4$ and $d=10$ meters. 
Thus, the approximation with the specific number $M$ of points is accurate. The slope of the expected harvested energy  for the baseline (linear)
schemes is different compared to the exact model, demonstrating their mismatch compared to the reality. 

In Fig.~\ref{fig:EhE_vs_Pt}-Right,
using the same small- and large-scale fading parameters as above,      $M+1 = 221$  approximation points, and distance   $d = 3$ m, it is shown that the linear model
is highly inaccurate for the second harvesting circuit module; thus, the widely adopted linear model cannot capture realistic efficiency models. The performance of the other
two baseline linear models is closer to the actual harvesting model. However, the slopes are different and a non-negligible  mismatch still exists.

Next, in Fig.~\ref{fig:EhE_vs_dist}-Left,
we depict the measured harvested power data from \cite{PowerCast} over the  input power range $[-15,-5]$ dBm, as well as the fitted harvested power functions 
obtained from:
(a)   the proposed model in~\eqref{eq:harvested_power_realistic_function} and 
(b)  the two nonlinear   models proposed in \cite{BoNgZlSc:15, XuOzAyMcKVi:17}.
For the  nonlinear  models of prior art,    
the normalized sigmoid function    \cite[Eqs.~(4) and~(5)]{BoNgZlSc:15} and the second-order polynomial 
in milliWatt scale    \cite[Eq.~(5)]{XuOzAyMcKVi:17} are utilized.
 The optimal parameters of the fitted functions  are obtained using the  Matlab's fitting toolbox.
It  can remarked that the proposed ground-truth harvested power model in Eq.~\eqref{eq:harvested_power_realistic}  fits perfectly to the measured data.
 The curve obtained using the sigmoid function in \cite{BoNgZlSc:15} tends to overestimate the measured harvested power for the small values of input power,
 while  the second-order polynomial   in \cite{XuOzAyMcKVi:17} underestimates the  harvested power for  the input power near sensitivity, offering 
 negative harvested power values for input power less than $-10$ dBm.

In Fig.~\ref{fig:EhE_vs_dist}-Right we depict the expected harvested energy 
as a function of distance using $P_{\rm T} = 2$ Watt comparing the above harvested power models.
 The path-loss model of Eq.~\eqref{path_loss_model}  is employed with $\nu = 2.1$ and $\mathtt{m}=5$.
The proposed piecewise linear approximation in Eq.~\eqref{eq:P_harv_approx} interpolates directly the measured $M+1 = 53$ data points  
 without  using any fitting.  
The harvested power model in  \cite{BoNgZlSc:15} overestimates the  expected harvested energy
for     large distances, deviating quite much from the 
reality. This stems directly from the fact that    the   sensitivity effects of the harvester are ignored in that model.
On the other hand,  the performance of  the model in  \cite{XuOzAyMcKVi:17} tends to underestimate the
expected harvested energy,  attaining   negative values for $d > 3.5$.
Compared to \cite{BoNgZlSc:15}, the model in  \cite{XuOzAyMcKVi:17} offers more accurate expected harvested energy performance for $d\leq 3.5$.
The proposed piecewise linear approximation, interpolating directly the measured harvested power data, 
achieves the same performance with the exact model.

\subsection{Time-Switching RF Energy Harvesting Scenario: Expected Charging Time}

 \begin{figure}[!t]
 \centering
         \includegraphics[width=0.6\columnwidth]{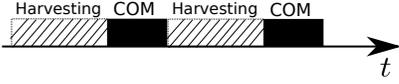}
  \caption{Time-switching operation. Necessary energy is harvested before the communication, in duty-cycled, non-continuous applications (e.g., wireless sensors).}
 \label{fig:time_switching}
 \end{figure}

Another important metric is the expected time for the 
RF harvesting circuit to charge its storage unit at the minimum required level, before operation. This is graphically illustrated in Fig.~\ref{fig:time_switching}, showing the
\emph{time-switching} RF energy harvesting and communication protocols, where the terminal
(e.g., a wireless sensor) first scavenges the necessary energy for transmission and then 
communicates (e.g., work in \cite{AsDaBle:16}). This is typical in many RF harvesting protocols, 
since the available power density in $\mu$Watt/cm$^2$ is limited and cannot sustain the power requirements 
of the overall apparatus; thus, a duty-cycled, non-continuous operation is necessary, as depicted in Fig.~\ref{fig:time_switching}.
The time needed to harvest the necessary energy before operation should be accurately quantified.

An energy  harvesting outage event after $N$ coherence periods will occur if the harvested 
energy after $N$ coherence periods is below a threshold. The latter is determined by
the  capacity of the energy storage unit (e.g., a capacitor $\mathtt{C}$) and the operating voltage $\mathtt{V}$ of the harvesting circuit.
Thus, the outage event is given by:
\begin{align}
\mathscr{O}_N   \triangleq &~ \left\{   T_{\rm p} \, \sum_{n=1}^N     \pmb{P}_{\rm har}^{(n)}
\leq \frac{1}{2} \, \mathtt{C} \, \mathtt{V}^2  \right\} =
  \left\{   \sum_{n=1}^N \!   \pmb{P}_{\rm har}^{(n)} 
\leq  \theta_{\rm harv}^{\rm th}\right\},
\label{eq:energy_harvesting_outage_event}
\end{align}
where the power threshold is determined   by the 
minimum required stored energy for operation 
$\frac{1}{2} \, \mathtt{C} \, \mathtt{V}^2 $,  as well as 
the transmission duration $T_{\rm p}$, i.e.,
 $\theta_{\rm harv}^{\rm th} \triangleq  \frac{\mathtt{C} \, \mathtt{V}^2}{2 \,  T_{\rm p}}$.
  Note that the above event depends
on the fading coefficients $\{\pmb{\gamma}^{(n)}\}_{n \in [N]}$.

The RV   $\pmb{N}^{\star}$ is defined as the first coherence time index when the accumulated harvested power is above the power threshold $\theta_{\rm harv}^{\rm th}$, 
given that there exist $\pmb{N}^{\star}-1$ consecutive outage events; thus, the probability mass function (PMF) of RV $\pmb{N}^{\star}$ can be  derived as:
\begin{align}
&\mathbb{P}( \pmb{N}^{\star} = N) \nonumber \\
\triangleq & \mathbb{P} \left( \mathscr{O}_{N-1}  \cap 
\left\{   \pmb{P}_{\rm har}^{(N)}   > \theta_{\rm harv}^{\rm th} - \sum_{n=1}^{N-1} \!  
 \pmb{P}_{\rm har}^{(n)}  \right\} 
 \right) \nonumber \\
  = &  \mathbb{P} \left(   \sum_{n=1}^{N-1} \!   \pmb{P}_{\rm har}^{(n)} 
\leq  \theta_{\rm harv}^{\rm th}  \cap \pmb{P}_{\rm har}^{(N)}  > \theta_{\rm harv}^{\rm th} - \sum_{n=1}^{N-1} \!   \pmb{P}_{\rm har}^{(n)}  \right) \nonumber \\
\overset{(a)}{=} &  \mathbb{P}\!\left(   \pmb{U}_{N-1} \leq \theta_{\rm harv}^{\rm th} \cap  \pmb{U}_{N-1} >   \theta_{\rm harv}^{\rm th}  -\pmb{P}_{\rm har}^{(N)}   \right) \nonumber \\
\overset{(b)}{=} & \!\!  \int_{x \in \mathbf{dom} \mathsf{f}_{\pmb{P}_{\rm har}}} \!\!\!\! \!\!\!\! \!\!\! \mathbb{P}\!\left(   \pmb{U}_{N-1}
\leq \theta_{\rm harv}^{\rm th} \cap  \pmb{U}_{N-1} >   \theta_{\rm harv}^{\rm th} 
 -x \right)    \, \mathsf{f}_{\pmb{P}_{\rm har}^{(N)}}(x) \mathsf{d} x \nonumber \\
 \overset{(c)}{=} &  \mathsf{F}_{\pmb{U}_{N-1}}\!(\theta_{\rm harv}^{\rm th} ) - \!  \int_{x \in \mathbf{dom} \mathsf{f}_{\pmb{P}_{\rm har}}} \!\!\!\!
\mathsf{F}_{\pmb{U}_{N-1}}\!(\theta_{\rm harv}^{\rm th} - x )  \, \mathsf{f}_{\pmb{P}_{\rm har}^{(N)}}(x) \mathsf{d} x,
\label{eq:prob_N_star}
\end{align}
where step $(a)$ used the definition of RV  $\pmb{U}_{N}$, i.e., $ \pmb{U}_{N-1} =   \sum_{n=1}^{N-1} \!   \pmb{P}_{\rm har}^{(n)}$, 
 step $(b)$ exploited the law of iterated expectation and the fact that $\pmb{U}_{N-1}$ and $\pmb{P}_{\rm har}^{(N)}$ are independent, and
step $(c)$ employed the CDF definition. Note that the expression above requires the CDF of 
$\pmb{U}_{N-1}$, which will be offered subsequently, while PDF of $\pmb{P}_{\rm har}^{(N)}$ can be given with the methodology of Section~\ref{sec:statistical_descr_P_harv}.

The expected value of discrete RV $\pmb{N}^{\star }$
can be easily calculated as:
\begin{equation}
\mathbb{E}[\pmb{N}^{\star}] \triangleq \overline{N}^{\star} = \sum_{N=1}^{\infty}  N \cdot \mathbb{P}( \pmb{N}^{\star} = N).
\end{equation}
The physical meaning of $\overline{ {N}}^{\star} $ is the average number
of coherence periods, i.e.,  $\overline{{N}}^{\star} \, T_{\rm c}$  seconds,  required for the
capacitor charging, before the communication. Such expected charging time is a prerequisite time interval, necessary for 
scavenging adequate RF energy for any subsequent  operation.


A numerical methodology to calculate ${\overline{N}}^{\star}$  is provided  
for the proposed  approximation model in~\eqref{eq:P_harv_approx}.
To calculate ${\overline{N}}^{\star}$ for the proposed model, Eq.~\eqref{eq:prob_N_star} must be exploited
using   $\widetilde{ \pmb{U} }_{N-1} \triangleq   \sum_{n=1}^{N-1} \!   \widetilde{\pmb {P}}_{\rm har}^{(n)} $ and $\widetilde{\pmb {P}}_{\rm har}^{(N)} $.
However,  only the PDF of  each individual RV $\widetilde{\pmb {P}}_{\rm har}^{(n)}$, $n \in [N]$, is known.
Hence, a methodology to calculate the CDF and the PDF of $\widetilde{ \pmb{U} }_{N-1}$ is proposed, exploiting
the fact that the latter  can be written as a sum of independent RVs.
The proposed methodology to evaluate Eq.~\eqref{eq:prob_N_star}, and thus $\overline{N}^{\star}$, 
is provided in Appendix~\ref{app:desnity_evolution}.
Applying the  methodology  presented in Appendix~\ref{app:desnity_evolution}, 
the PMF  of RV $\pmb{N}^{\star}$ is calculated for the proposed model using Eq.~\eqref{approx_integral_PMF_N_star}
for any threshold $\theta_{\rm harv}^{\rm th}$.

 \begin{figure*}[!t]
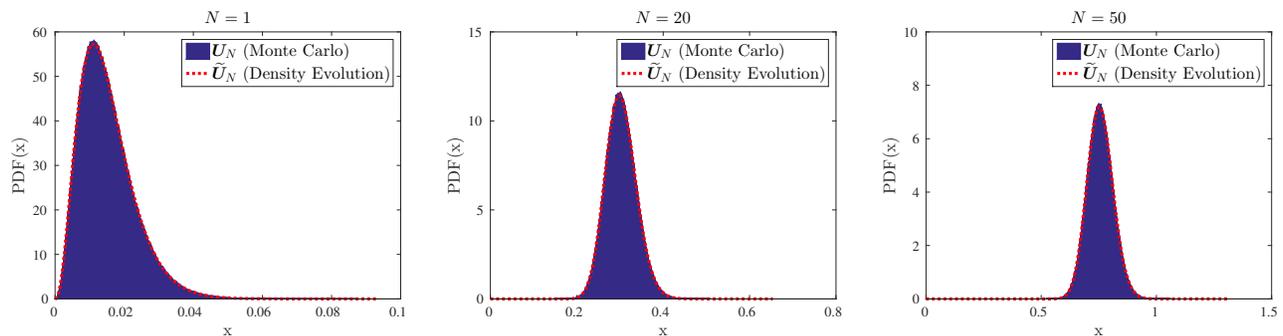

 \centering
         \includegraphics[width=0.6\columnwidth]{fig13.eps} \quad
  \includegraphics[width=0.6\columnwidth]{fig14.eps} \quad
  \includegraphics[width=0.6\columnwidth]{fig15.eps}
  \caption{The histogram of actual $\pmb{U}_N$ and the corresponding PDF vector $\mathbf{\underline{v}}_{\mathsf{f}}$
for $N=1$, $N=20$, and $N=50$ for the energy harvesting model in  \cite{AsDaBle:16}. }
 \label{fig:density_evolution_different_N}
 \end{figure*}


\begin{figure*}[!h]
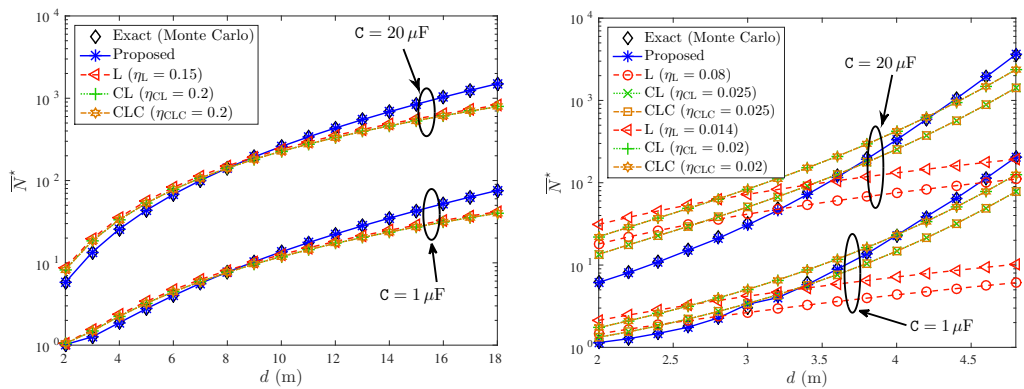

 \centering
         \includegraphics[width=0.75\columnwidth]{fig16.eps}  \quad 
 \includegraphics[width=0.72\columnwidth]{fig17.eps} 
  \caption{Left (Right): Expected number of coherence periods $\pmb{N}^{\star}$, $\overline{N}^{\star}$,
  necessary for charging vs. {distance} for the rectenna proposed in  \cite{AsDaBle:16} 
(PowerCast module  \cite{PowerCast}).}
 \label{fig:E_chag_time_vs_Ptx}
 \end{figure*}

Consider the rectenna model in \cite{AsDaBle:16}, the  path-loss model given in~\eqref{path_loss_model} with $\nu = 2.1$ and  $d = 5$ m, transmission power
$P_{\rm T}= 1.5 $ Watt, Nakagami parameter $\mathtt{m} = 5$, while the parameters for the power threshold
are set to $\mathtt{V} = 1.8$ V,  $\mathtt{C} = 10$ $\mu$F,  $T_{\rm p} = 50$ msec. 
Fig.~\ref{fig:density_evolution_different_N} shows the histogram
of actual $\pmb{U}_N$ and the corresponding estimated PDF of RV $\widetilde{\pmb{U}}_N$,
for $N=1$, $N=20$, and $N=50$.\footnote{Appendix~\ref{app:desnity_evolution} parameters are  $H=2^{16} $,  $I_{\rm Lo}= 0$, $I_{\rm Up} = N \mathbb{E}\!\left[
 \widetilde{ P}_{\rm har}^{(n)}\right]+ 10 \sqrt{N \,  \mathsf{var}\!\left[ \widetilde{ P}_{\rm har}^{(n)}\right]}$,
 $\mathtt{G} = \frac{I_{\rm Up} - I_{\rm Lo} }{H}$,  and $J_{\rm FFT} = 2^{17}$.}
It can be seen that the red dotted curves 
corresponding to the estimated PDFs, and the actual PDF (histogram) 
are perfectly matched.

 %

\subsubsection{Numerical Results}

Fig.~\ref{fig:E_chag_time_vs_Ptx} depicts
the expected $N^{\star}$ for the realistic, proposed, and baseline models  
as a function of distance for different capacitor values  ($\mathtt{C} = 1$ $\mu$F and $\mathtt{C} = 20$ $\mu$F) 
for the two harvesting efficiency models  in  \cite{AsDaBle:16} (Left)
and    \cite{PowerCast} (Right) using $\mathtt{V} = 1.8$ V and  $T_{\rm p} = 50$ msec. 
The path-loss model in Eq.~\eqref{path_loss_model} is employed for the evaluation in conjunction with Nakagami fading. 
In Fig.~\ref{fig:E_chag_time_vs_Ptx}-Left (Right) the utilized wireless channel parameters are $\nu = 2.1$,   $\mathtt{m} = 5$,
$P_{\rm T} = 1.5$ Watt,  while for the density evolution, the following parameters are employed $H = 2^{17}$ and $J_{\rm FFT} = 2^{18}$ 
($J_{\rm FFT} = 2^{19}$).
The number of data points to approximate the  harvested power in Eq.~\eqref{eq:P_harv_approx}
was $M+1 = 1171$ and $M+1 = 2201$ data points for the rectennas in  \cite{AsDaBle:16}  
and   \cite{PowerCast}, respectively.

For both  harvesting efficiency models in \cite{AsDaBle:16} and  \cite{PowerCast} the  
expected charging time  for the proposed  approximation and the true, nonlinear harvested power model
coincide, corroborating the accuracy of a) the proposed approximation in Eq.~\eqref{eq:P_harv_approx} and b) the framework in Appendix~\ref{app:desnity_evolution}.

For the baseline models, the results are obtained through Monte Carlo. It is observed that 
although the results for baseline models are offered with the best possible values for $\eta_{\rm L}$, $\eta_{\rm CL}$, and $\eta_{\rm CLC}$,
the baseline linear harvesting efficiency models fail to offer the same slope with the true, nonlinear  energy harvesting  model; as a result, the obtained $N^{\star}$  {for} the linear models may deviate one order of magnitude from the true value, offering consequently deviations from the true duty-cycle and the available resources for wireless communications. It is also noted that the presence
of a boost converter at the rectifier output may also magnify the necessary time for charging, further amplifying the charging time differences. 
The proposed methodology with the nonlinear harvesting model is clearly able to offer accurate estimation of the charging time.

\subsection{Power-Splitting RF Energy Harvesting Scenario: Passive RFID Tags}

  \begin{figure}[!b]
 \centering
         \includegraphics[width=0.6\columnwidth]{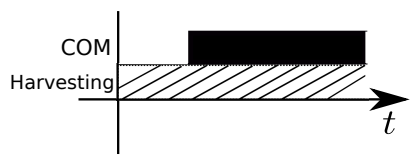}
  \caption{The power-splitting operation  mode.}
 \label{fig:power_splitting}
 \end{figure}

Next,  a  backscatter RFID scenario is considered where the EIH node is a passive RFID tag that splits the input RF power for operation 
and wireless communication, simultaneously (Fig.~\ref{fig:power_splitting}), as opposed to the time-switching (duty-cycled) operation. 
The passive RFID tags typically use a simple RF switch (e.g., a transistor) to communicate with an  interrogator.

 \begin{figure}[!t]
 \centering
 \includegraphics[width=0.7\columnwidth]{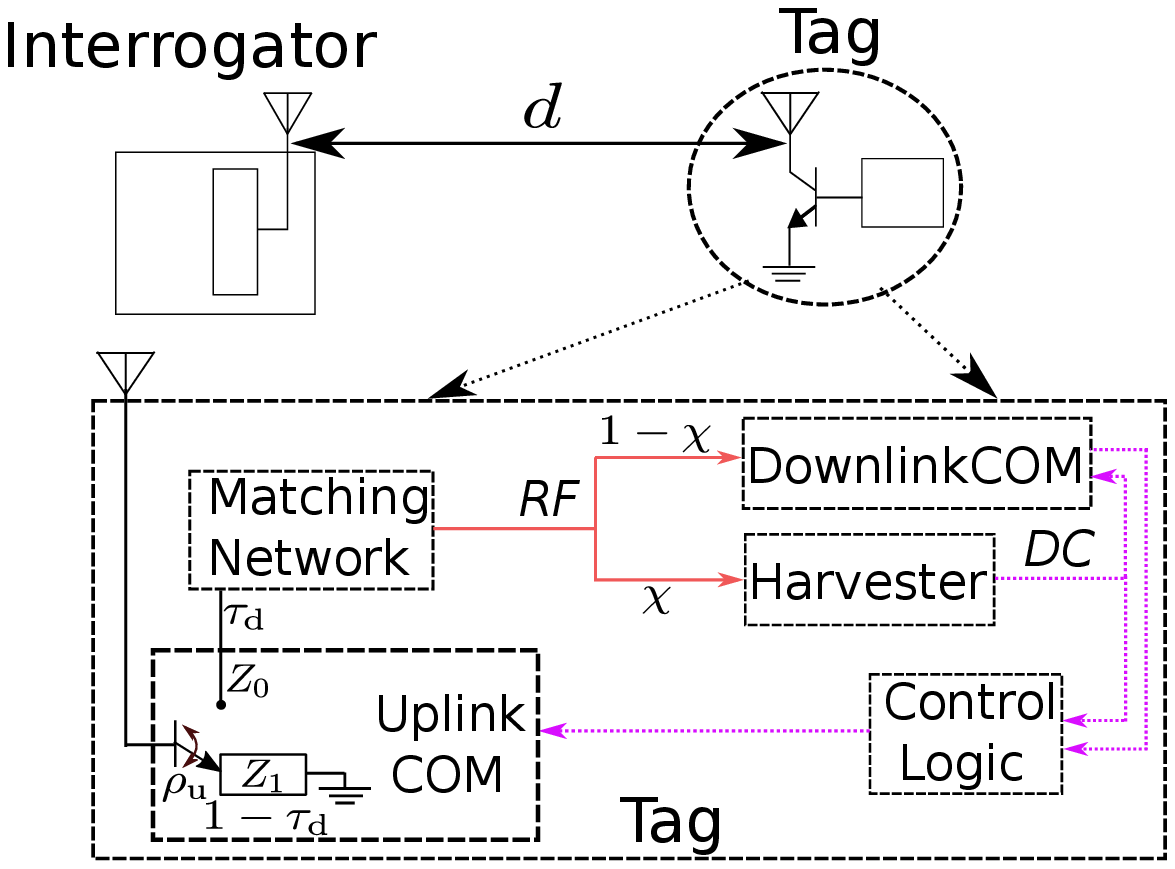} 
  \caption{A monostatic backscatter architecture consisting of an interrogator  (i.e., an RFID reader)
  and a passive RFID tag. The interrogator's antenna acts as the 
  transmitter of illuminating signal, as well as the receiver of reflected, i.e., the backscattered (from tag) information, hence the term \emph{monostatic}.}
 \label{fig:reader_tag_RFID}
 \end{figure}

A typical operating block  diagram  of a passive RFID tag  is depicted in Fig.~\ref{fig:reader_tag_RFID}.
Suppose that the tag's  antenna is terminated between two load values $Z_0$ and $Z_1$.
When the antenna is terminated at $Z_0$, it is matched to input load and  the tag absorbs the power from the incident
signal. When the antenna is terminated
at load $Z_1$, the tag reflects the incoming signal, i.e., it scatters back information (uplink), provided that it has sufficient amount of energy. 
It is further assumed that the overall round-trip communication
among the interrogator and the tag  lasts a single coherence time period, thus
we focus on a single coherence time block; thereinafter,    coherence block index $n$ is removed to simplify the notation.

Parameter $\tau_{\rm d}$ denotes the fraction of time  the antenna load is
at $Z_0$ (absorbing state), while the rest $1- \tau_{\rm d}$ corresponds to fraction of time at load
$Z_1$ (reflection state). Assume that $\chi$ is the fraction of the   input power (when tag's antenna
load is at absorbing state) dedicated for the RF energy harvesting operation; thus, a total of $\zeta_{\rm har} = \chi \,\tau_{\rm d} $ percentage of the input power is
dedicated for energy harvesting, with $\zeta_{\rm har}  \in (0,1)$. 
The rest  $ (1- \chi )\tau_{\rm d}$ input signal power is exploited by the tag downlink communication circuitry. Furthermore, 
a  fraction $\rho_{\rm u} \leq  1- \tau_{\rm d}$ of the impinged power 
is used for the uplink scatter radio operation.  This number
depends on the scattering efficiency and the fraction of time the tag's antenna is terminated at the load $Z_1$. 
It is noted that the scattering efficiency depends on the reflection coefficients, which in turn are input power-independent.
With monostatic architecture, the incident input power at tag is $\pmb{P}_{\rm R} =P_{\rm T } \, \mathsf{L}(d) \, \pmb{\gamma}  =\mathsf{P}(d) \pmb{\gamma}$.
Since, only a fraction  $\rho_{\rm u}$ of the input power is backscattered  (i.e., $\rho_{\rm u} \, \pmb{P}_{\rm R} $),
 the received power at the interrogator  due to the round trip nature of   backscattering operation is 
\begin{equation}
\mathsf{g}_{\rm int}(\pmb{P}_{\rm R}) \triangleq \rho_{\rm u} \,  \pmb{P}_{\rm R}  \,  \mathsf{L}(d) \,\gamma 
 = \rho_{\rm u} \frac{( \pmb{P}_{\rm R})^2}{ P_{\rm T }}.
\label{eq:P_received_interrogator}
\end{equation}

The two following events are needed:
\begin{align}
\mathscr{A} & \triangleq \{ \text{The BER at the interrogator is below a threshold $\beta$} \}   \nonumber \\
&= \left\{  2 \,   \mathsf{Q}\!\left(\frac{\sqrt{\mathsf{g}_{\rm int}(\pmb{P}_{\rm R}) }}{\sigma_{\rm u}}\right)\left (1 -  
\mathsf{Q}\!\left(\frac{ \sqrt{\mathsf{g}_{\rm int}(\pmb{P}_{\rm R}) }}{\sigma_{\rm u}}\right) 
 \right) < \beta  \right \}
\label{eq:BER_event}
\end{align}
and
\begin{align}
\mathscr{B} & \triangleq \{ \text{\small The harvested power  is larger than tags' power consumption $\mathtt{P}_{\rm c}$} \}   \nonumber \\
&= \left\{    \mathsf{p}( \zeta_{\rm har}  \,  \pmb{P}_{\rm R} )  > \mathtt{P}_{\rm c}  \right \},
\label{eq:energy_outage_event}
\end{align}
where $\mathsf{Q}(x) =\frac{1}{\sqrt{2 \pi}} \int_x^{\infty} \mathsf{e}^{-\frac{t^2}{2}} \mathsf{d}t $ is the  Q-function
 and the expression in the last line of Eq.~\eqref{eq:BER_event} is the probability of bit error under coherent   
maximum-likelihood detection with FM$0$ line coding \cite{KaMaBl:15}, {and  $\beta \in  (0, \frac{1}{2})$ is the BER threshold}. 
Parameter $\sigma_{\rm u}^2$ is a properly scaled variance of thermal AWGN noise at the receiving circuit of the 
interrogator. The expression in~\eqref{eq:BER_event} can be further simplified with the aid of the following:
\begin{proposition} \normalfont
The function 
\begin{equation}
y = \mathsf{R}(x) \triangleq    2 \,\mathsf{Q}( x ) \, (1 -  \mathsf{Q}(x))  , ~x \in  (0,\infty),
\label{eq:R_function}
\end{equation}
is   monotone decreasing  and invertible over the positive reals;
the  inverse function is  given by  
\begin{equation}
x = \mathsf{R}^{-1}(y) =  \mathsf{Q}^{-1}\!\left( \frac{1 - \sqrt{1 - 2\,y}}{2} \right)  ,   ~y \in (0,0.5),
\label{eq:R_inv_function}
\end{equation}  
where the  function $ \mathsf{Q}^{-1}(\cdot)$ denotes the inverse of Q-function (with respect to composition). 
\begin{proof}
The proof  is given in Appendix~\ref{sec:proof_prop_inv_R_function}.	
\end{proof}
\label{prop:inverse_R_function}
\end{proposition}


 \begin{figure*}[!t]
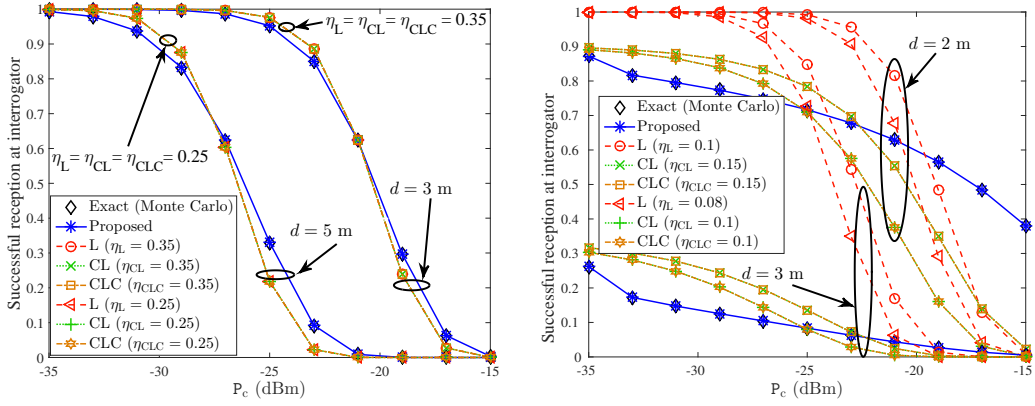

 \centering
         \includegraphics[width=0.75\columnwidth]{fig20.eps} \quad
         \includegraphics[width=0.75\columnwidth]{fig21.eps}
  \caption{Left (Right):  The probability of successful reception at interrogator  as a function of tags' power consumption $\mathtt{P}_{\rm c}$ and tag-interrogator distance, 
for the rectenna model  in  \cite{AsDaBle:16} (PowerCast module \cite{PowerCast}).}
 \label{fig:RFID_prob_succ_reception_vs_Pc}
 \end{figure*}

The event of the successful interrogator reception is denoted by $\mathscr{S}$; 
the non-successful reception event  at the interrogator, $\mathscr{S}^{\rm C}$, occurs
if a) the harvested power is below the tag's power consumption or  (b) given that the
harvested power is above the tag's power consumption $\mathtt{P}_{\rm c}$, the BER at the interrogator is above the threshold $\beta$:
\begin{align}
\mathbb{P}(\mathscr{S}^{\rm C}) &= \mathbb{P}( \mathscr{B}^{\rm C}) +  \mathbb{P}(\mathscr{A}^{\rm C} | \mathscr{B}) \mathbb{P}(\mathscr{B})
=1 -  \mathbb{P}( \mathscr{B}) +  \mathbb{P}(\mathscr{A}^{\rm C} | \mathscr{B}) \mathbb{P}(\mathscr{B}) \nonumber \\
&= 1 - \mathbb{P}( \mathscr{B})(1 - \mathbb{P}(\mathscr{A}^{\rm C} | \mathscr{B})) =
1 - \mathbb{P}( \mathscr{B})\mathbb{P}(\mathscr{A} | \mathscr{B}) \nonumber \\
 &= 1 -  \mathbb{P}(\mathscr{A} \cap \mathscr{B})  =  1 - \mathbb{P}(\mathscr{S}).
\label{eq:error_event_interrogator}
\end{align}
Thus, in view of Eq.~\eqref{eq:error_event_interrogator}, the probability of successful event is expressed as:
\begin{align}
\mathbb{P}(\mathscr{S}) =& ~\mathbb{P} \left(      \mathsf{R}\!\left(\frac{\sqrt{\mathsf{g}_{\rm int}(\pmb{P}_{\rm R}) }}{\sigma_{\rm u}} \right) < \beta  \cap
\mathsf{p}( \zeta_{\rm har}  \,  \pmb{P}_{\rm R} ) > \mathtt{P}_{\rm c}      \right) \nonumber \\
\overset{(a)}= & ~
\mathbb{P}\! \left(     \pmb{P}_{\rm R}  > \frac{ \sqrt{P_{\rm T}} \,  \mathsf{R}^{-1}(\beta) \sigma_{\rm u} }{\sqrt{\rho_{\rm u} }} \cap
  \mathsf{p}( \zeta_{\rm har}  \,  \pmb{P}_{\rm R} ) >  \mathtt{P}_{\rm c}       \right),
\label{eq:suc_reception_reader}
\end{align}
  where in step $(a)$ we exploited the fact that the  function $\mathsf{R}^{-1}$ in~\eqref{eq:R_inv_function} is monotone decreasing 
and then we  plugged  the definition of function $\mathsf{g}_{\rm int}(\cdot)$.

 The corresponding probability expressions  can be derived for  the baseline linear models and the proposed nonlinear harvesting model. 
The successful reception event at the interrogator  for 
baseline models is denoted as $\mathscr{S}_{\rm c}$, $\rm c \in \{ L,CL,CLC\}$
and for the proposed model as $\widetilde{\mathscr{S}}$. The following proposition summarizes the results:
\begin{proposition} \normalfont
Suppose that $\mathtt{P}_{\rm c} > 0$ and consider Nakagami fading. 
Let us define threshold $\theta_{\mathscr{A}}  \triangleq \frac{ \sqrt{P_{\rm T}} \,  \mathsf{R}^{-1}(\beta) \sigma_{\rm u} }{\sqrt{\rho_{\rm u} }} > 0$.  
For the linear model, the probability of event $\mathscr{S}_{\rm L}$
is given by: 
\begin{equation}
\mathbb{P}(\mathscr{S}_{\rm L}) = \frac{\sGamma\!  \left(  \mathtt{m},  \frac{ \mathtt{m}}{   \mathsf{P}(d)} \theta_{\rm max}^{\rm L} \right)}{\Gamma(\mathtt{m})} ,
\end{equation}
where $ \theta_{\rm max}^{\rm L} \triangleq \max\{ \theta_{\mathscr{A}} , \frac{\mathsf{p}_{\rm L}^{-1}(\mathtt{P}_{\rm c})}{\zeta_{\rm har}}\}$.

For the constant-linear model, the probability of event $\mathscr{S}_{\rm C L}$
is given by: 
\begin{equation}
\mathbb{P}(\mathscr{S}_{\rm CL}) = \frac{\sGamma\!  \left(  \mathtt{m},  \frac{ \mathtt{m}}{
 \mathsf{P}(d)} \theta_{\rm max}^{\rm CL} \right)}{\Gamma(\mathtt{m})} ,
\end{equation}
where $ \theta_{\rm max}^{\rm CL} \triangleq \max\{\theta_{\mathscr{A}} , \frac{\mathsf{p}_{\rm CL}^{-1}(\mathtt{P}_{\rm c})}{\zeta_{\rm har}}\}$.

For the last baseline model (CLC), the   probability of event $\mathscr{S}_{\rm C LC}$
is expressed as follows:
\begin{equation}
\mathbb{P}(\mathscr{S}_{\rm CLC}) =
\begin{cases}
 \frac{\sGamma  \left(  \mathtt{m},  \frac{ \mathtt{m}}{
 \mathsf{P}(d)} \theta_{\rm max}^{\rm CLC} \right)}{\Gamma(\mathtt{m})} ,&  0<\mathtt{P}_{\rm c} < \mathsf{p}_{\rm CLC}(\mathtt{P}_{\rm in}^{\rm sat}),  \\
0, & \mathtt{P}_{\rm c} \geq \mathsf{p}_{\rm CLC}(\mathtt{P}_{\rm in}^{\rm sat}),
\end{cases}
\end{equation}
where $ \theta_{\rm max}^{\rm CLC}\triangleq \max\{\theta_{\mathscr{A}} , \frac{\mathsf{p}_{\rm CLC}^{-1}(\mathtt{P}_{\rm c})}{\zeta_{\rm har}}\}$.

Finally, for the proposed nonlinear energy harvesting model,   the   probability of event $\widetilde{\mathscr{S}}$
is given by:
\begin{equation}
\mathbb{P}(\widetilde{\mathscr{S}}) = 
\begin{cases}
    \frac{\sGamma\!  \left(  \mathtt{m},  \frac{ \mathtt{m}}{
 \mathsf{P}(d)} \widetilde{\theta}_{\rm max} \right)}{\Gamma(\mathtt{m})}
,  & 0<\mathtt{P}_{\rm c} < v_M, \\
0 , &  \mathtt{P}_{\rm c} \geq v_M,
\end{cases}
\label{eq:reception_interr_proposed_model}
\end{equation}
where $ \widetilde{\theta}_{\rm max}\triangleq \max\{\theta_{\mathscr{A}} , \frac{\widetilde{\mathsf{p}}^{-1}(\mathtt{P}_{\rm c})}{\zeta_{\rm har}}\}$.
\begin{proof}
The proof  can be found  in  Appendix~\ref{sec:proof_theorem_rec_interrogator}.	
\end{proof}
\label{prop:reception_at_interrogator}
\end{proposition}

\subsubsection{Numerical Results}

Fig.~\ref{fig:RFID_prob_succ_reception_vs_Pc} offers the probability of successful reception
at the interrogator, as function of  the tag power consumption $\mathtt{P}_{\rm c}$ and the tag-interrogator distance
under the path-loss model of Eq.~\eqref{path_loss_model}. The following parameters are utilized: $\tau_{\rm d} = 0.5$,
$\chi = 0.5$, $\rho_{\rm u}= 0.01$, $\beta = 10^{-5}$, $\sigma_{\rm u}^2 = 10^{-11}$ mWatt.
In Fig.~\ref{fig:RFID_prob_succ_reception_vs_Pc}-Left (Right) the rectenna model in \cite{AsDaBle:16} (harvesting module in \cite{PowerCast})
is studied using parameters $\nu = 2.1$, $\mathtt{m}= 5$,  $P_{\rm T}= 1.5$ Watt ($P_{\rm T}= 3$ Watt), under 
two distance setups:  $d = 5$ m and $d = 3$ m ($d = 3$ m and $d = 2$ m), and using     $M+1 = 586$   ($M+1 = 221$)   data points.

From both figures it can be seen that  the performance of the proposed  approximation  in Eq.~\eqref{eq:P_harv_approx} 
is the same with the performance of the real model in   Eq.~\eqref{eq:harvested_power_realistic}. 
On the other hand, the baseline models offer different slopes compared to the nonlinear model and 
fail to approach its  performance; this holds for both harvesting circuits, even though deviations are more obvious for the harvester in \cite{PowerCast}; 
  it is also noted  that the selected values of $\eta_{\rm L}$, $\eta_{\rm CL}$, and $\eta_{\rm CLC}$ were chosen so as to reduce the performance difference.
It is worth noting  that the linear model's performance curve has completely different slope and curvature
compared to the real   model.
Again, it can be deduced that the proposed harvesting model and the offered methodology provide  accurate results in sharp contrast to the linear harvesting models.

\section{Conclusions}
\label{conclusion}
 
For the first time in the RF energy harvesting
 literature, realistic efficiency models are studied
 accounting for the sensitivity, nonlinearity, and saturation
 of the RF harvesting circuits. 
  The impact of harvester's sensitivity is carefully quantified. 
 A piecewise linear approximation
 model is proposed, amenable to closed-form, tuning-free modeling, and expressions. Using two real   rectenna 
models from   RF harvesting circuits' prior art, it is demonstrated that the proposed 
approximation model is in complete agreement with   reality, whereas
  linear or nonlinear-infinite sensitivity  RF harvesting modeling results deviate
from the reality.
  It is deduced  that  the SWIPT research should take into account the nonlinearity of the actual harvesting efficiency and the 
  limited sensitivity of the harvester.
 
\section*{Acknowledgment}
The authors would like to thank Georgios Vougioukas for the brainstorming and the proofreading of the manuscript.

\appendices

\section{Proof of Proposition~\ref{prop:PDF_CDF_Pharv}}
\label{sec:proof_prop_CDF_Pharv}

Here   the CDF expression in Eq.~\eqref{eq:CDF_P_harv} is shown.
Using the PDF of Eq.~\eqref{eq:PDF_P_harv},   for  any
$x \in [v_{m-1} ,v_m] \backslash \{ v_M\}$, $m \in [M]$:
\begin{align}
& \mathsf{F}_{\widetilde{\pmb {P}}_{\rm har}^{(n)}}(x) = \int_{0}^x  \mathsf{f}_{\widetilde{\pmb {P}}_{\rm har}^{(n)}}(y)  \mathsf{d}y   \nonumber \\
\overset{(a)} =&~ \sum_{j = 1}^{m-1}  \int_{v_{j-1}}^{v_j}   \frac{1}{l_j} \, \mathsf{f}_{\pmb{P}_{\rm R}^{(n)} }\!\! \left( 
\frac{y  -  v_{j-1} + l_j b_{j-1}}{l_j} \right) \mathsf{d}y + \nonumber \\
+  &~ \int_{v_{m-1}}^{x}   \frac{1}{l_m} \, \mathsf{f}_{\pmb{P}_{\rm R}^{(n)} }\!\! \left( 
\frac{y  -  v_{m-1} + l_m b_{m-1}}{l_m} \right) \mathsf{d}y   \nonumber \\
\overset{(b)} = &~  \sum_{j = 1}^{m-1}  \int_{b_{j-1}}^{b_j}   \mathsf{f}_{\pmb{P}_{\rm R}^{(n)} }\!(y) \mathsf{d}y
+ \int_{b_{m-1}}^{\frac{x  -  v_{m-1} + l_m b_{m-1}}{l_m}}   \mathsf{f}_{\pmb{P}_{\rm R}^{(n)} }\!(y) \mathsf{d}y \nonumber \\
\overset{ }{= }  &  
\mathsf{F}_{\pmb{P}_{\rm R}^{(n)} }\! \left(\frac{x  -  v_{m-1} + l_m b_{m-1}}{l_m} \right),
\label{eq:CDF_proof_x_less_vM}
\end{align}
where in $(a)$,  the integral is divided in a
sum of  integrals associated with  disjoint intervals and
in $(b)$,    change of variables $y' =\frac{y  -  v_{j-1} + l_j b_{j-1}}{l_j}$ is performed
for each individual integral.
Note that due to the right-continuity of the CDF  \cite{pappoulis:02}, Eq.~\eqref{eq:CDF_proof_x_less_vM}
covers the case of $x = v_0 = 0$ since $ \mathsf{F}_{\widetilde{\pmb {P}}_{\rm har}^{(n)}}(0) =  \mathsf{F}_{\pmb{P}_{\rm R}^{(n)} }\! \left(\frac{ l_1 b_{0}}{l_1} \right)
= \xi_0$.

For  $x \geq  v_M$, the following holds 
\begin{align}
 \mathsf{F}_{\widetilde{\pmb {P}}_{\rm har}^{(n)}}(x)  \overset{(a)}{=} &  \int_{0}^{v_M^-}
 \mathsf{f}_{\widetilde{\pmb {P}}_{\rm har}^{(n)}}(y)  \mathsf{d}y  + 
 \int_{v_M}^{x} \mathsf{f}_{\widetilde{\pmb {P}}_{\rm har}^{(n)}}(y)  \mathsf{d}y   \nonumber \\
\overset{(b)}{=}  & \,  \xi_M +  (1- \xi_M) = 1,
\label{eq:CDF_proof_x_geq_vM}
\end{align}
where in $(a)$,  the integral  is divided
over the disjoint intervals $[0 ,v_M) $ and $[v_M, x)$, while in $ (b)$,
we plugged the definition of 
 the  CDF   found in   Eq.~\eqref{eq:CDF_proof_x_less_vM} over   interval $[0 ,v_M)$,
and we used the definition of PDF in~\eqref{eq:PDF_P_harv} for  $x \geq  v_M$.
The above conclude the proof.

\section{Proof of Proposition~\ref{prop:approximation_error_poly}}
\label{sec:proof_prop_approx_err}
The proof of this proposition relies on \cite[Th.~6.2]{SuMay:03}. 
For any  continuously differentiable function $\mathsf{g}(\cdot)$  defined over 
an interval $ [x_0,x_1]$ and a linear function $\widetilde{\mathsf{g}}(\cdot)$ that interpolates $\mathsf{g}(\cdot)$ 
on $x_0$ and $x_1$, for any $x \in [x_0,x_1] $ there exists $\phi \equiv \phi(x) \in (x_0,x_1) $  satisfying the following
\begin{equation}
\mathsf{g}(x) -  \widetilde{\mathsf{g}}(x) = \frac{(x-x_0)(x-x_1)}{2}  \mathsf{g}''(\phi),
\label{eq:interpolation_theorem}
\end{equation}
 where $\mathsf{g}''(\cdot)$ denotes the second
order derivative of function $\mathsf{g}(\cdot)$. Using   Eq.~\eqref{eq:interpolation_theorem},
 the absolute error is upper bounded as
\begin{align}
 \int_{x_0}^{x_1} \!   \left | \mathsf{g}(x) \! - \! \widetilde{\mathsf{g}}(x) \right|  \mathsf{d}x & \! \leq 
\! \frac{1}{2}\max_{x \in [x_0,x_1]} \! | \mathsf{g}''(x) |\! \!  \int_{x_0}^{x_1} \! \!\! \!\!\left| (x-x_0)(x-x_1)  \right| \mathsf{d}x    \nonumber \\
 & \! = \!
\frac{1}{2} \, \mathtt{C}_{\mathsf{g}} \!  \int_{x_0}^{x_1} \!(x-x_0)(x_1 - x)   \mathsf{d}x   ,
\label{eq:_integral_error_general_1}
\end{align}
where the constant $\mathtt{C}_{\mathsf{g}} \equiv \mathtt{C}_{\mathsf{g}}(x_0,x_1) 
\triangleq   \max_{x \in [x_0,x_1]}  | \mathsf{g}''(x) |$
depends on function $\mathsf{g}(\cdot)$, as well as the 
points $x_0$ and $x_1$.
Combining the following identity 
\begin{equation}
\max_{x \in [x_0, x_1] } (x-x_0)(x_1-x) = \frac{(x_1-x_0)^2}{4}
\label{eq:upper_bound_quadratic}
\end{equation}
with   Eq.~\eqref{eq:_integral_error_general_1}, 
the absolute error can be upper bounded   as
\begin{equation}
 \int_{x_0}^{x_1}  \left | \mathsf{g}(x) -  \widetilde{\mathsf{g}}(x) \right|  \mathsf{d}x \leq  \frac{\mathtt{C}_{\mathsf{g}} (x_1-x_0)^3 }{8}.
\label{eq:_integral_error_general_2}
\end{equation}

Next, the above framework is  applied to the proposed piecewise linear   function $\widetilde{\mathsf{p}}(\cdot)$.
Since $\mathsf{p}(\cdot)$ is continuously differentiable  in  $\mathcal{P}_{\rm in}$,
using the fact that  $\mathsf{p}(b_m) = \widetilde{\mathsf{p}}(b_m)$, for $m=0,1,\ldots, M$, and
applying the results above, the following is obtained
\begin{align}
 & \int_{\mathcal{P}_{\rm in}}     \left | \mathsf{p}(x) -  \widetilde{\mathsf{p}}(x) \right|  \mathsf{d}x 
=  \sum_{m=1}^{M }\int_{b_{m-1}}^{b_{m} }     \left | \mathsf{p}(x) -  \widetilde{\mathsf{p}}(x) \right|  \mathsf{d}x 
\nonumber \\
\overset{(a)}{\leq} &  \frac{(\delta_M)^3}{8} \sum_{m=1}^{M } \max_{x \in [b_{m-1},b_{m}] }  \left | \mathsf{p}''(x) \right| \nonumber \\
\overset{(b)}{\leq} &  \frac{(\delta_M)^3}{8}   M   \max_{x \in \mathcal{P}_{\rm in}}\left| \mathsf{p}''(x) \right|=
\frac{  \mathtt{C}_{\mathsf{p}} \, (\mathtt{P}_{\rm in}^{\rm sat} - \mathtt{P}_{\rm in}^{\rm sen})^3}{8 \, M^2}.
\end{align}
where in $ (a)$,   $\delta_M = b_{m} - b_{m-1} $ is utilized, combined
 with the result in~\eqref{eq:_integral_error_general_2}, while in 
$(b)$,   $ \max_{x \in \mathcal{P}_{\rm in}}\left | \mathsf{p}''(x) \right| \geq 
\max_{x \in [b_{m-1},b_{m}] }  \left | \mathsf{p}''(x) \right|$ for any $m \in [M]$ is employed.
Constant  $\mathtt{C}_{\mathsf{p}} \equiv \mathtt{C}_{\mathsf{p}} (\mathcal{P}_{\rm in}) \triangleq 
  \max_{x \in \mathcal{P}_{\rm in}}\left | \mathsf{p}''(x) \right|$
depends on   set  $ \mathcal{P}_{\rm in}$ and the given function  $\mathsf{p}(\cdot)$,
and is independent of $M$.

\section{Numerical Density Evolution Framework for the  Sum of Independent RVs}
\label{app:desnity_evolution}

Consider a RV $\pmb{x}$ which is expressed as $\pmb{x} = \sum_{n=1}^{N} \pmb{x}^{(n)}$, where  RVs $\{\pmb{x}^{(n)}\}_{n=1}^{N}$ are independent of each other,
supported by       sets $\mathcal{S}^{(n)}$, $n \in [N]$, respectively. It is assumed that the PDF of each individual
RV $\pmb{x}^{(n)}$, $\mathsf{f}_{\pmb{x}^{(n)}}(\cdot)$, is given  over the support $\mathcal{S}^{(n)}$, $n\in [N]$,
and each $\mathcal{S}^{(n)}$ is bounded.
 In addition note that the support  of the RV 
$\pmb{x}$ is $\mathcal{S} = \mathcal{S}^{(1)} + \mathcal{S}^{(2)} + \ldots + \mathcal{S}^{(N)}$ (set addition),
due to the required convolution operation.

The idea of density evolution is to approximate numerically the PDF of 
RV $x $ exploiting the fact that it can be written as the convolution of
individual PDFs. To do so,
consider the support set $[I_{\rm Lo},  I_{\rm Up}] $ as an approximation of
set $\bigcup_{n=1}^N \mathcal{S}^{(n)} \cup \mathcal{S}$.
Note that set can be chosen so as $ \int_{y \in  [I_{\rm Lo},  I_{\rm Up}]} \mathsf{f}_{\pmb{x}^{(n)}}(y) \mathsf{d}y \approx 1$, $\forall n \in [N]$,
and $\int_{y \in  [I_{\rm Lo},  I_{\rm Up}]} \mathsf{f}_{\pmb{x}}(y) \mathsf{d}y \approx 1$.
The support set   $[I_{\rm Lo},  I_{\rm Up}]$ is discretized  using $H+1$ grid
points with uniform grid resolution $\mathtt{G} = \frac{ I_{\rm Up} - I_{\rm Lo}}{H}$, and the following
discrete (support) set is formed  
\begin{equation}
\mathcal{H}_{\mathtt{G}}  = \{ I_{\rm Lo}   +  h\,\mathtt{G}  \}_{h = 0}^{H}.
\label{eq:support_set}
\end{equation} 
Set $\mathcal{H}_{\mathtt{G}}$ is a discrete approximation of  support $ [I_{\rm Lo},  I_{\rm Up}]$
and  can be also viewed as a vector
with $H+1$ elements, whose the $j$-th element is $ \mathcal{H}_{\mathtt{G}} [j] = I_{\rm Lo}   +  (j-1)\mathtt{G}$.
Let us denote  $\mathbf{\underline{v}}_{\mathsf{f}}^{(1)}, \mathbf{\underline{v}}_{\mathsf{f}}^{(2)} , \ldots,  \mathbf{\underline{v}}_{\mathsf{f}}^{(N)}$
 the  $H+1$-dimensional  PDF vector representations of RVs $\pmb{x}^{(1)}, \pmb{x}^{(2)}, \ldots, \pmb{x}^{(N)}$, respectively, 
where each element of $\mathbf{\underline{v}}_{\mathsf{f}}^{(n)}$ is given by
\begin{equation}
\mathbf{\underline{v}}_{\mathsf{f}}^{(n)}[j]   \triangleq \mathsf{f}_{\pmb{x}^{(n)}}(\mathcal{H}_{\mathtt{G}} [j]),~ j  \in [ H+1] . 
\label{eq:PDFxl_vec_vl}
\end{equation}
Note that  with the above definition of PDF vector $\mathbf{\underline{v}}_{\mathsf{f}}^{(n)} $, the following approximation holds:
  $1  = \int_{y \in \mathcal{S}^{(n)}} \mathsf{f}_{\pmb{x}^{(n)}}(y) \mathsf{d}y \approx \sum_{j=1}^{H+1} \mathbf{\underline{v}}_{\mathsf{f}}^{(n)}[j] \,\mathtt{G} $,
 for each $n \in [N]$.

Next, using $J_{\rm FFT} > H + 1$ points (for efficient implementation
$J_{\rm FFT}$ has to be a power of 2)   the fast Fourier transform (FFT)
of PDF $\mathbf{\underline{v}}_{\mathsf{f}}^{(n)}$ is evaluated, which is the characteristic function of RV $\pmb{x}^{(n)}$.
The vector of  the characteristic function of the RV $\pmb{x}^{(n)}$ is given by
\begin{equation}
 \mathbf{\underline{r}}^{(n)} = \mathsf{FFT}\!\left( \widetilde{ \mathbf{\underline{v}}}_{\mathsf{f}}^{(n)}  \,\mathtt{G} \right) \in  \mathds{C}^{J_{\rm FFT}}
\end{equation}
where $(\widetilde{ \mathbf{\underline{v}}}_{\mathsf{f}}^{(n)})^{\top} = \left[(\mathbf{\underline{v}}_{\mathsf{f}}^{(n)})^{\top}~~ \mathbf{\underline{0}}_{J_{\rm FFT} - (H + 1)}^{\top}  \right]^{\top}$ 
is the zero-padded version of $ \mathbf{\underline{v}}_{\mathsf{f}}^{(n)}$, appending extra   {$J_{\rm FFT} - (H + 1)$}
zeros  at the end   of $ \mathbf{\underline{v}}_{\mathsf{f}}^{(n)}$.
Using the following facts: (a) the sum of independent RVs is the convolution of their associated PDFs and
 (b) the equivalence among convolution operation and the inverse Fourier transform of the product of the Fourier transforms,
 the final PDF of $\pmb{x}$ is obtained as
 \begin{equation}
 \mathbf{\underline{v}}_{\mathsf{f}_{\pmb{x}}}=  \mathsf{IFFT}\!\left(\mathbf{\underline{r}}^{(1)} \odot \mathbf{\underline{r}}^{(2)} \odot \ldots  \odot  \mathbf{\underline{r}}^{(N)}\right)[1:H+1] 
\label{eq:PDFx_vec_v}
\end{equation} 
where vector $ \mathbf{\underline{v}}_{\mathsf{f}_{\pmb{x}}}$ consists of the first $H+1$ elements of the vector 
$\mathsf{IFFT}(\mathbf{\underline{r}}^{(1)} \odot \mathbf{\underline{r}}^{(2)} \odot \ldots  \odot  \mathbf{\underline{r}}^{(N)})$
and   is an approximation of the PDF of RV $\pmb{x}$.
The CDF vector   representation  for   RV $\pmb{x}$ can be   evaluated as
\begin{equation}
\mathbf{\underline{v}}_{\mathsf{F}_{\pmb{x}}}[j] = \sum_{i=1}^{j} \mathbf{\underline{v}}_{\mathsf{f}_{\pmb{x}}}[i] \,\mathtt{G}, ~j\in [H+1].
\label{eq:CDFx_vec_v}
\end{equation}
Note with the above methodology the evaluation 
of  $\mathbf{\underline{v}}_{\mathsf{f}_{\pmb{x}}}$ requires only $\mathcal{O}(N \,J_{\rm FFT}\, \mathsf{log} J_{\rm FFT} )$
arithmetic operations due to the properties of FFT \cite{GolVan:89}. 

To evaluate Eq.~\eqref{eq:prob_N_star} for a given threshold $\theta$,
  the PDF of RV $\pmb{u} = \sum_{n=1}^{N-1} \pmb{x}^{(n)}$, $\mathbf{\underline{v}}_{\mathsf{f}_u}$, is first calculated   using
  Eq.~\eqref{eq:PDFx_vec_v} with $N-1$. Then, the index associated with largest  element of $\mathcal{H}_{ \mathtt{G}}$ that is smaller than $\theta$ is found,
i.e., if $\theta^*  =\arg \max \{  y \in  \mathcal{H}_{ \mathtt{G}}: y  \leq \theta \} $  the optimal
index $j_{\theta}$ satisfies $\theta^* = \mathcal{H}_{ \mathtt{G}}[j_{\theta}]$, 
and then   we calculate the discrete approximation of~\eqref{eq:prob_N_star} as 
\begin{equation}
\mathbf{\underline{v}}_{\mathsf{F}_{\pmb{u}}}[j_{\theta}] - \sum_{i=1}^{j_{\theta}  } \mathbf{\underline{v}}_{\mathsf{F}_{\pmb{u}}}[j_{\theta} - i +1] \, 
\mathbf{\underline{v}}^{(N)}_{\mathsf{f}}[i] \,\mathtt{G}.
\label{approx_integral_PMF_N_star}
\end{equation}
The overall complexity to calculate $\overline{N}^{\star}$ for the proposed model
is dominated by the calculation of  $\mathbf{\underline{v}}_{\mathsf{f}_{\pmb{u}}}$
which is  $\mathcal{O}(N \,J_{\rm FFT}\, \mathsf{log} J_{\rm FFT} )$.
 
\section{Proof of Proposition~\ref{prop:inverse_R_function}}
\label{sec:proof_prop_inv_R_function}

 By differentiating Eq.~\eqref{eq:R_function} with respect to $x$, after some basic algebra,
 we obtain for $x > 0$ 
 \begin{equation}
  \mathsf{R}'(x)   =  2 \mathsf{Q}'(x)(1 - 2\mathsf{Q}(x)) \!\overset{(a)}{=}\! \frac{-2 \mathsf{e}^{-\frac{x^2}{2}}}{2  \pi} (1 -  2\mathsf{Q}(x)) \! \overset{(b)}{<} \!0, 
 \end{equation}
where in $(a)$, we plugged the derivative of function $\mathsf{Q}(\cdot)$, i.e., $ \mathsf{Q}'(x) =  \frac{-2 \mathsf{e}^{-\frac{x^2}{2}}}{2  \pi} $,
while in  $(b) $,   $\mathsf{Q}(x) < 0.5$ for every $x >0$ was used. Since   $\mathsf{R}'(x) <0$, for $x > 0$,
the function $\mathsf{R}(\cdot)$ is monotone decreasing, and thus, invertible in $(0,\infty)$.
Since $y = \mathsf{R}(x) \in \left(0,\frac{1}{2}\right)$ for $x \in (0,\infty)$,
solving the equation $y = 2\mathsf{Q}(x) - 2\mathsf{Q}^2(x)$, the valid answer
is $\mathsf{Q}(x) = \frac{1 - \sqrt{1 - 2y}}{2} \in  \left(0,\frac{1}{2}\right)$.
Therefore, since $\mathsf{Q}(\cdot)$ is a monotone function, the inverse of $\mathsf{R}(\cdot)$ becomes
\begin{equation}
x =  
\mathsf{Q}^{-1}\!\left(\frac{1- \sqrt{1-2y}}{2}\right), ~y \in \left(0,\frac{1}{2} \right).
\end{equation}

\section{Proof of Proposition~\ref{prop:reception_at_interrogator}}
\label{sec:proof_theorem_rec_interrogator}

 The proof is provided for the proposed model,
as the rest baseline models are   special cases.
The proof for the baseline models can be obtained using similar reasoning.
First note that since the image points are selected as $0 < v_1 < v_2 < \ldots < v_M$, the slopes satisfy
$l_1 < l_2 < \ldots < l_M$; thus,  the
piecewise linear function $\widetilde{\mathsf{p}}(\cdot)$
is monotone increasing in $[b_0, b_M] $ (and thus, invertible in $[0,v_M]$).

Firstly, consider the case $0<\mathtt{P}_{\rm c}< v_M$, implying that $ b_0 < \widetilde{\mathsf{p}}^{-1}(\mathtt{P}_{\rm c}) < b_M$.
Using similar reasoning with Eq.~\eqref{eq:suc_reception_reader},
 the probability of successful reception at interrogator for the proposed model can be expressed as
\begin{align}
\mathbb{P}(\widetilde{\mathscr{S}}) \triangleq &~
\mathbb{P}\! \left(   \pmb{P}_{\rm R}  > \frac{ \sqrt{P_{\rm T}} \,  \mathsf{R}^{-1}(\beta) \sigma_{\rm u} }{\sqrt{\rho_{\rm u} }} \cap
  \widetilde{\mathsf{p}}(\zeta_{\rm har} \pmb{P}_{\rm R}) >  \mathtt{P}_{\rm c}    \right) \nonumber \\
\overset{(a)}{=}  & ~\mathbb{P}\! \left(    \pmb{P}_{\rm R} >  \theta_{\mathscr{A}}  \cap
   \pmb{P}_{\rm R} > \frac{\widetilde{\mathsf{p}}^{-1}(\mathtt{P}_{\rm c})}{\zeta_{\rm har}}    \right)   
\overset{(b)}{=}    \mathbb{P}\! \left(  \pmb{P}_{\rm R} > \widetilde{\theta}_{\rm max}  \right) \nonumber \\
\overset{ }{=} & ~
1 - \mathsf{F}_{   \pmb{P}_{\rm R}} \!(\widetilde{\theta}_{\rm max} ),
\end{align}
where $(a)$ stems from the definition of $\theta_{\mathscr{A}}$
as well as the fact that $0<\mathtt{P}_{\rm c}< v_M$, while $(b)$ relies on the definition of $\widetilde{\theta}_{\rm max} $.
The result follows by plugging the CDF of  $\pmb{P}_{\rm R}$
for Nakagami fading.

For $\mathtt{P}_{\rm c} \geq v_M$, the following  holds 
\begin{equation}
\widetilde{\mathscr{S}} \!\subseteq\!
\left\{  \widetilde{\mathsf{p}}(\zeta_{\rm har} \pmb{P}_{\rm R}) >  \mathtt{P}_{\rm c}  \right \}
\!\overset{(a)}{\subseteq} \!
\left \{  \widetilde{\mathsf{p}}(  \pmb{P}_{\rm R}) >  v_M  \right \}  \!= \!\left \{   \widetilde{ \pmb{P} }_{\rm har}  >  v_M \right \}  , 
\label{eq:event_S_proposed}
\end{equation}
where $(a)$ results from the following  facts:  (i) $ \mathtt{P}_{\rm c} \geq v_M$ and
(ii) $\widetilde{\mathsf{p}}(\zeta_{\rm har} \pmb{P}_{\rm R}) \leq \widetilde{\mathsf{p}}(  \pmb{P}_{\rm R})$,
since  $\zeta_{\rm har} \in (0,1)$  and the function $\widetilde{\mathsf{p}}(\cdot)$ is non-decreasing. 
Thus,  by the monotonicity of probability measure \cite{Fol:99}, Eq.~\eqref{eq:event_S_proposed} implies that 
$ \mathbb{P}( \widetilde{\mathscr{S}}) \leq  \mathbb{P}\!\left( \widetilde{ \pmb{P} }_{\rm har}  >  v_M \right)
= 1 - \mathsf{F}_{ \widetilde{ \pmb{P} }_{\rm har}}\!(v_M) = 0$; the last equality holds due to
the definition of CDF in  Eq.~\eqref{eq:CDF_P_harv}.
Hence,   for $\mathtt{P}_{\rm c} \geq v_M$,  $\mathbb{P}(\widetilde{\mathscr{S}}) = 0$.

\balance


\end{document}